\documentclass[letterpaper,showkeys,showpacs,amsmath,amssymb,pre,groupedaddress,12pt,preprint,tightenlines]{revtex4}

\usepackage{amsfonts}
\usepackage{graphicx}
\usepackage{natbib}
\usepackage{dcolumn}
\usepackage{bm}
\usepackage[letterpaper,left=0.8in,right=0.8in,top=0.8in,bottom=0.8in]{geometry}
\usepackage{rotating}
\usepackage{longtable}

\begin{document}

\title{The International-Trade Network:\\Gravity Equations and Topological Properties}

\author{Giorgio Fagiolo}

\affiliation{Sant'Anna School of Advanced Studies\\Laboratory of Economics and
Management\\Piazza Martiri della Libert\`{a} 33\\I-56127 Pisa, Italy\\ Tel:
+39-050-883356 Fax: +39-050-883343\\Email: giorgio.fagiolo@sssup.it}

\date{August 2009}

\begin{abstract}
\noindent This paper begins to explore the determinants of the
topological properties of the international - trade network (ITN).
We fit bilateral-trade flows using a standard gravity equation to
build a ``residual'' ITN where trade-link weights are depurated from
geographical distance, size, border effects, trade agreements, and
so on. We then compare the topological properties of the original
and residual ITNs. We find that the residual ITN displays, unlike
the original one, marked signatures of a complex system, and is
characterized by a very different topological architecture. Whereas
the original ITN is geographically clustered and organized around a
few large-sized hubs, the residual ITN displays many small-sized but
trade-oriented countries that, independently of their geographical
position, either play the role of local hubs or attract large and
rich countries in relatively complex trade-interaction patterns.



\end{abstract}

\maketitle

\section{Introduction}
The last years have witnessed the emergence of a large body of
contributions addressing international-trade issues from a
complex-network perspective
\citep{LiC03,SeBo03,Garla2004,Garla2005,ReichardtWhite2007,serrc07,Bhatta2007a,Bhatta2007b,
Garla2007,Tzekina2008,Fagiolo2008physa,Fagiolo2008acs,Fagiolo2009pre}.
The International Trade Network (ITN), aka World-Trade Web (WTW) or
World-Trade Network (WTN), is defined as the graph of import/export
relationships between world countries in a given year.

Understanding the topological properties of the ITN, and their evolution over
time, acquires a fundamental importance in explaining issues such as economic
globalization and internationalization, the spreading of international crises,
and the transmission of economic shocks, for two related reasons
\citep{HelliwellPadmore1985,Artis2003,Forbes2002}. On the one hand, direct
(bilateral) trade linkages are known to be one of the most important channels
of interaction between world countries \citep{Krugman1995}. On the other hand,
they can only explain a small fraction of the impact that an economic shock
originating in a given country can have on another one, which is not among its
direct-trade partners \citep{AbeFor2005}. Therefore, a complex-network analysis
\citep{AlbertBarabasi2002,DoroMendes2003} of the ITN, by characterizing in
detail the topological structure of the network, can go far beyond the scope of
standard international-trade indicators (such as ``openness to trade''
\footnote{Traditionally measured by the ratio of exports plus imports to
country's gross domestic product (GDP).}), which instead only account for
bilateral-trade direct linkages \citep{Fagiolo2008acs}. Trade paths connecting
any pair of non-direct trade partners may then shed light on the likelihood
that economic shocks might be transmitted between the two countries
\citep{KaliReyes2007}, and possibly help explaining macroeconomic dynamics
\citep{Kali_etal_2007}.

The first stream of studies that have explored the properties of the
ITN has employed a \textit{binary-network analysis}, where a
(possibly directed) link between any two countries is either present
or not according to whether the value of the associated trade flow
is larger than a given threshold \citep{SeBo03,Garla2004,Garla2005}.
More recently, however, a growing number of contributions
\citep{LiC03,Bhatta2007a,Bhatta2007b,Garla2007,Fagiolo2008physa,Fagiolo2009pre}
have adopted a \textit{weighted-network approach}
\citep{Barrat2004pnas,Barthelemy2005} to the study of the ITN.
There, a link between any two countries is weighted by the
(deflated) value of trade (imports plus exports) that has occurred
between these countries in a given time interval. A set of
weighted-network topological measures \citep{Barthelemy2005} is then
computed to characterize the architecture of the weighted ITN.

The motivation for a weighted-network analysis is that a binary approach cannot
fully extract the wealth of information about the trade intensity flowing
through each link and therefore might dramatically underestimate the role of
heterogeneity in trade linkages. Interestingly, Refs.
\citep{Fagiolo2008physa,Fagiolo2009pre} show that the statistical properties of
the ITN viewed as a weighted network crucially differ from those exhibited by
its weighted counterpart, and that a weighted-network analysis is able to
provide a more complete and truthful picture of the ITN than a binary one.

Notwithstanding much is known about the topological properties of
the ITN ---in both its binary and weighted versions--- and how they
have evolved in the recent past, a set of fundamental questions
remains to be answered: What are the determinants of such
properties? Are there relevant node or link characteristics (other
than the ones related to trade flows) that can explain the peculiar
topological patterns actually observed in the ITN? Such questions
might be in principle addressed from a theoretical perspective, i.e.
looking for models of network formation and evolution that have as
their equilibria graphs with properties similar to those actually
observed in the ITN (cf. for example Refs.
\citep{AlbertBarabasi2002,Jackson03survey}). In this paper we take
an applied approach and attempt to explore the foregoing issues from
a more empirical perspective.

To begin with, note that all network-related topological variables are
univocally obtained from the link-weight distribution \footnote{For example,
node strength, clustering coefficient, centrality, etc. are simple
manipulations that require as inputs the link-weight distribution only. See
Appendix C for formal definitions.}. Hence, if (as typically happens) one
weights the link between country $i$ and country $j$ in a given year by the sum
of the deflated values of imports of $i$ from $j$ and exports of $i$ to $j$ in
that year, then all topological properties of the weighted ITN defined in that
way will entirely depend on the matrix of international bilateral-trade flows
observed in the year under study, which plays the role of sufficient
statistics. This is straightforward for the weighted-version of the ITN, but is
also true for its binary representations, as the probability that a given link
is present still depends, given the chosen threshold, on the distribution of
observed bilateral-trade flows. As a result, one might safely conclude that
much of what we know about the topological properties of the ITN can be
empirically accounted for by the set of statistically-significant independent
variables that explain international bilateral-trade flows.

In the context of the ITN literature, this issue has been initially
addressed by Refs. \citep{Garla2004,Garla2007}, who have shown that
the probability that any two countries are connected, as well as the
value of the trade flow between them, is well explained by (the
product of) their current GDP, which plays the role of a ``hidden''
node variable or fitness. More generally, we know from the huge
empirical literature on gravity equations
\citep{LeamerLevinsohn1995,Overman2003} that international-trade
flows can be almost entirely explained by a multiplicative model
featuring as independent variables the GDPs of the two countries
involved in the trade link, their geographical distance (as a proxy
of all factors that might create trade resistance, e.g. transport
costs), and a series of dummies accounting for other geographical,
social, historical and political factors (e.g., existence of common
borders, religion and languages, colonial ties, trade agreements,
and so on). By fitting a gravity equation to the original
international bilateral-trade data one may then account for
explanatory variables of link weights.

Such an exercise might be interesting for two related reasons.
First, from a gravity-equation approach, one may identify what are
the main determinants of international-trade flows in the data used
to build trade networks. Second, and more importantly, one might
think to remove all the existing structure from the data to check
whether the residual weighted ITN exhibits topological features
comparable to those of the original ITN.

This paper explores these two lines of inquiry by performing the
following simple exercise. In line with the recent literature
\citep{LiC03,Bhatta2007a,Bhatta2007b,Garla2007,Fagiolo2008physa,Fagiolo2009pre},
we start from a weighted-network representation of the ITN in a
given year where a link between any two countries is weighted by the
(deflated) value of their total trade (import plus exports). We then
fit total-trade flows, i.e. the entries of the ITN weight matrix,
with a standard gravity model. This allows us to identify the
relative impact of size, geographical distance, and other
geographical, social, historical and political factors, on the
weighted-network representation of the ITN under study. Finally, we
build a ``residual'' version of the ITN where each link is now
weighted by the associated residual of the fitted gravity model.
This allows us to remove much of the structure that is conceivably
present in the original data.

We ask whether ---and to what extent--- the topological properties
of such a residual ITN mimic those of the original version of the
ITN (as studied, e.g., in Ref. \citep{Fagiolo2007pre}). Notice that
a similar approach has been already used in Ref.
\citep{Fagiolo2009jee}. They find that, by and large, the ITN
architecture remains unaltered if ones removes the GDP dependence
only from link weights (e.g., if one employs as link weights the
ratio between bilateral-trade flows divided by the GDP of
\textit{either} the importer \textit{or} the exporter country). The
gravity-equation fit allows us to generalize this approach and
obtain a weighted ITN that depends on underlying factors, either
unobserved or not accounted for in the regression, related e.g. to
country technological similarities, degree of specialization, etc..

Our results show that the residual ITN is characterized by power-law
shaped distributions of link weights and node statistics (e.g.,
strength, clustering, random-walk betweenness centrality). Hence,
the underlying architecture of the weighted ITN seem to display
signatures of complexity. This must be contrasted with the original
ITN, where log-normal distributions were ubiquitous. We also find
that the correlation structure among node statistics, and between
node statistics and country per-capita GDP, changes substantially
when comparing the original and residual weighted networks. Whereas
the original ITN is geographically clustered and organized around a
few large-sized hubs, the residual ITN features many small-sized but
trade-oriented countries that, independently of their geographical
position, either play the role of local hubs or attract large and
rich countries in relatively complex trade-interaction patterns.

The rest of the paper is organized as follows. Section
\ref{Sec:Data} briefly presents the data sets employed in the paper.
Section \ref{Sec:Gravity} contains the results of gravity-equation
fits. A comparison of the topological properties of the original and
the ``residual'' version of the ITN is carried out in Section
\ref{Sec:Results}. Finally, Section \ref{Sec:Conclusions} concludes
discussing the implications of our results for modeling and
suggesting extensions of the present work.

\section{Data and Definitions}\label{Sec:Data}

We employ international-trade data provided by Ref.
\cite{GledData2002} to build a time-sequence of weighted directed
ITNs. Our balanced panel refers to $T=20$ years (1981-2000) and
$N=159$ countries (see Appendix A for the list of acronyms and
countries in the panel). For each country and year, data report
directed trade flows (e.g., exports) in current US dollars, which we
properly deflate. Weight matrices are built following the flow of
goods. This means that rows represent exporting countries, whereas
columns stand for importing countries. Following Refs.
\citep{LiC03,Bhatta2007a,Bhatta2007b,Garla2007}, we begin by
initially defining the weight $\tilde{w}_{ij}^t$ of a link from $i$
to $j$ in year $t$ as total exports from $i$ to $j$. However, a
simple statistical analysis of weighted matrices suggests that
$\tilde{W}^t$ are sufficiently symmetric to justify an undirected
analysis for all $t$ \citep{Fagiolo2009pre}. We therefore symmetrize
the network by defining the entries of the new weight matrix $W^t$
as the arithmetic average of import and export flows, i.e.
$w_{ij}^t=\frac{1}{2}[\tilde{w}_{ij}^t+\tilde{w}_{ji}^t]$
\footnote{Following Ref. \citep{Baldwin2006}, we have also
replicated our exercises by using a geometric average of bilateral
flows without noticing relevant differences in the results.}.
Finally, in order to have $w_{ij}^t\in[0,1]$ (and to remove all
trend-related factors), for all $(i,j)$ and $t$, we re-normalize all
entries in $W^t$ by their maximum value
$w^t_{\ast}=max_{i,j=1}^{N}\{w_{ij}^t\}$. This means that the
symmetrized weight $w_{ij}^t$ is proportional to total trade
(imports plus exports) flowing through the link $ij$ in a given
year.

Note that Ref. \citep{Fagiolo2009pre} has shown that all topological
properties of the ITN have remained fairly stable in the period
1981-2000. Therefore, in what follows, we shall focus on year 2000
only for the sake of exposition (all results robustly hold in other
years). Our relevant ITN data is then the weight $N\times N$ matrix
$W=W^{2000}$, which is symmetric by construction and features only
zeroes in its main diagonal.

For each country in the panel, we also gather a long list of
variables traditionally employed in gravity-equation exercises, see
Appendix B for labels, sources and explanations. These variables can
be grouped in three classes. First, there are variables related to
trade resistance factors, as the geographical distance between
countries and the degree of country remoteness \citep{Bhavnani2002}.
Second, country-size effects are controlled by country GDP,
population and geographical area. Third, a number of country
variables or link-related dummies control for geographical (common
border, landlocking, continent), economic (trade agreements,
exchange rates, consumer price indices) and social/ political/
historical effects (common languages and religion, trade agreements,
common currency). Together, these factors have been shown to
successfully explain, in a way or in the other, international-trade
flows in gravity-equation econometric exercises. Therefore, in the
next section, we shall employ a standard gravity-equation setup to
explain the entries of the weight matrix $W$ characterizing the ITN
graph in year 2000.

\section{Fitting Gravity Equations to ITN Data}\label{Sec:Gravity}

The gravity equation has become the workhorse setup to study the
determinants of bilateral international-trade flows
\citep{Overman2003,Fratianni2009}. Its basic symmetric original
formulation, inspired by Newton's gravity equation, states that
total trade between any two countries in the world is directly
proportional to the product of country masses (e.g., their GDP) and
inversely proportional to their geographic distance \footnote{This
empirically-inspired law has been found to be consistent with a
number of theoretical foundations, e.g. trade specialization models,
monopolistic-competition frameworks with intra-industry trade,
Hecksher-Ohlin models, etc.; see, e.g., Refs.
\citep{Anderson1979,Bergstrand1985,Deardorff1998,Anderson2003}.}.

A strong gravity-like dependence of bilateral-trade flows on
geography and size is evident in our data. For example, Table
\ref{Tab:Within_Between} reports trade-flow shares within and
between macro geographical areas. It is easy to see that areas that
trade more, mostly trade within the same area, or with
geographically-close areas. Conversely, those that trade less, tend
to trade with less distant areas and/or with areas where there are
countries historically and culturally tied.

\begin{center}
--- TABLE \ref{Tab:Within_Between} ABOUT HERE ---
\end{center}

\begin{figure}
\begin{center}
\begin{minipage}[t]{7cm}
\includegraphics[width=7cm]{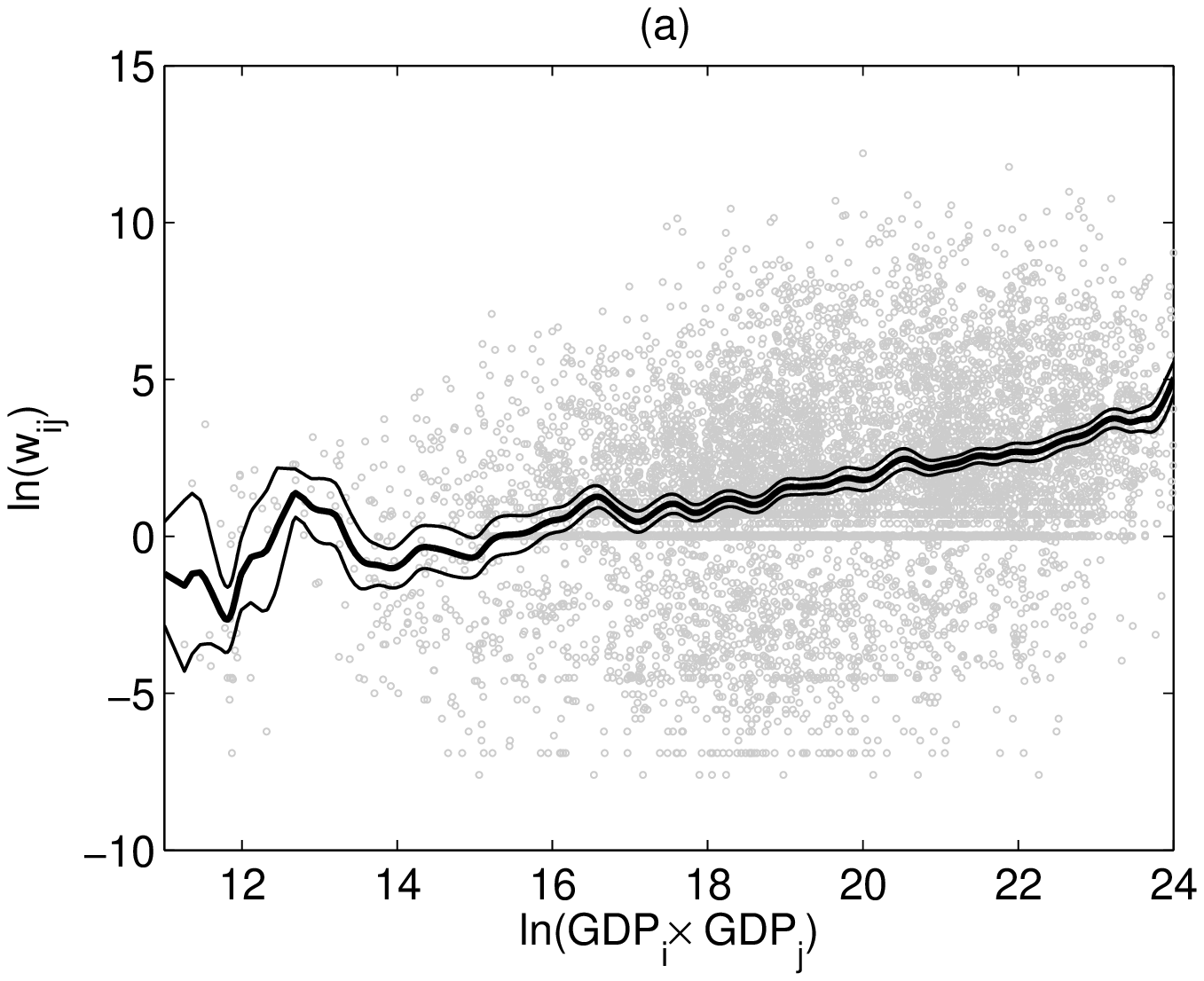}
\end{minipage}
\begin{minipage}[t]{7cm}
\includegraphics[width=7cm]{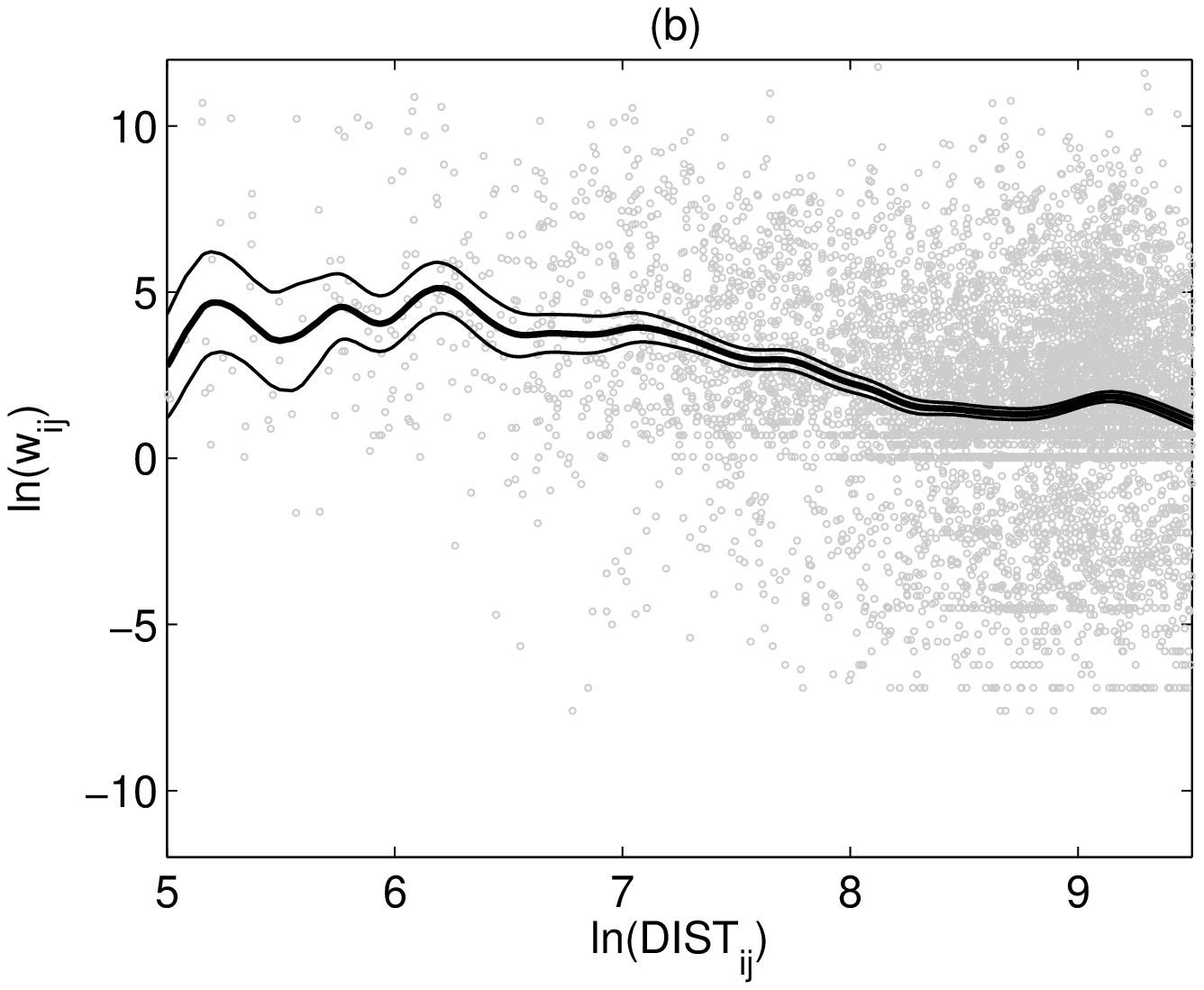}
\end{minipage}
\begin{minipage}[t]{7cm}
\includegraphics[width=7cm]{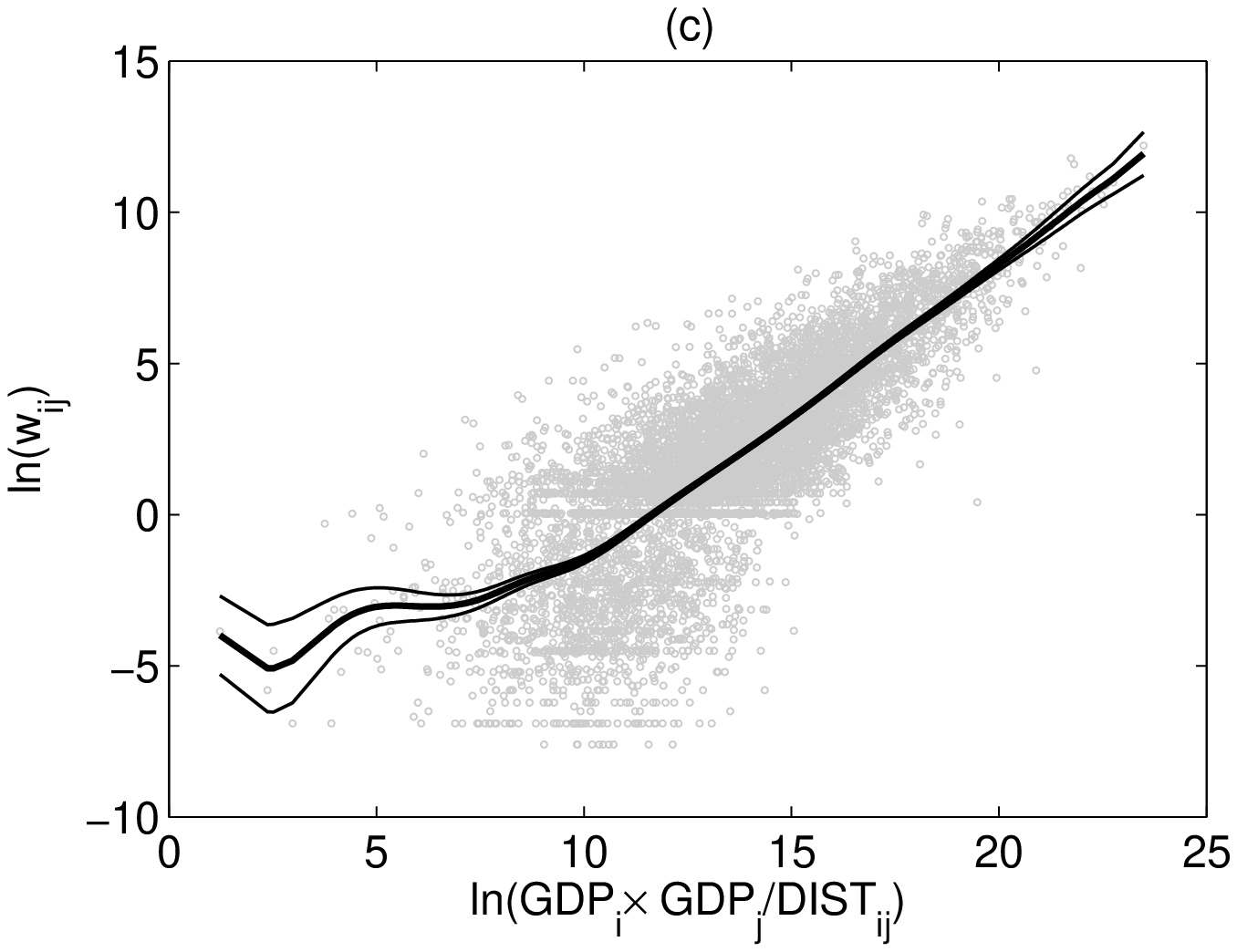}
\end{minipage}
\end{center}
\caption{Bilateral-trade flows vs. (a) product of country GDPs; (b) geographical distance; (c) product of country GDPs divided by geographical distance. Solid curves: Non-parametric estimate of bilateral-trade flow conditional mean with confidence bounds at 95\%.}
\label{Fig:Gravity_Scatters}
\end{figure}

Furthermore, a gravity-law structure clearly emerges if one studies
how bilateral trade flows $w_{ij}$ correlate with the product of
$i$'s and $j$'s GDPs and their geographical distance. Figure
\ref{Fig:Gravity_Scatters} plots in a log-log scale $w_{ij}$ against
$M_{ij}=GDP_i\cdot GDP_j$, $DIST_{ij}$ and $M_{ij}/DIST_{ij}$.
Non-parametric estimates of $w_{ij}$ conditional means, superimposed
on the clouds of points together with \%95 confidence bounds,
clearly indicate that both the correlation structure and the
functional form linking $w_{ij}$ to size and geographical effects
are in line with the prediction of the basic gravity law
\footnote{Conditional means have been estimated using the
local-linear non-parametric method proposed in Ref.
\citep{LiRacine2004}, employing cross-validated bandwidth selection
using the procedure of Ref. \citep{Hurvich_etal_1998}. Estimation
has been performed using the package {\tt np} \citep{np_package}
under R ({\tt http://www.r-project.org/}).}.

From an empirical perspective, the basic gravity equation has been
expanded in the literature to improve the fit by taking into account
country- or trade-specific characteristics that may influence
bilateral trade flows in addition to masses and distance (see
previous section). Overall, this strategy has proven itself to be
extremely successful in order to explore the determinants of
bilateral trade flows and we shall replicate it in what follows
\footnote{On this huge empirical literature, see for example Refs.
\citep{Overman2003,Frankel1997,RoseSpiegel2002,Rose2000,GlickRose2001,EichengreenIrwin1996,LeamerLevinsohn1995,Feenstra2001},
among others.}.

For our estimation purposes, it is convenient to start with the most general
multiplicative formulation of the gravity equation \citep{Santos2006}, which
reads:

\begin{eqnarray}
w_{ij}=\alpha_0 [GDP_i^{\alpha_1} GDP_j^{\alpha_2}DIST_{ij}^{\alpha_3}] [\prod_{k=1}^{K}{X_{ik}^{\beta_{ik}}X_{jk}^{\beta_{jk}}}] \cdot \nonumber \\ \label{Eq:Gravity} \\ \cdot exp\{\sum_{h=1}^{H}{\theta_{ijh}D_{ijh}} + \sum_{l=1}^{L}{\delta_{il}Z_{il}} + \gamma_{i}C_{i} + \gamma_{j}C_{j} \} \eta_{ij}. \nonumber
\end{eqnarray}
Here $w_{ij}, i,j=1,\dots,N$ (with $i>j$) is the
$N(N-1)/2$-dimensional vector of symmetric bilateral trade flows
(i.e. the upper-diagonal entries of the weight matrix $W$);
$(\alpha, \beta, \theta, \delta, \gamma)$  are unknown (vectors of)
parameters to be estimated; $X_{i}$ are $K$- dimensional vectors of
country-specific continuous or categorical variables (other than
GDP); $Z_{i}$ are $L$-dimensional vectors of country-specific dummy
variables; $D_{ijh}$ are $H$-dimensional vectors of link-specific
dummy variables (other than distance); $C_{i}$ are country-specific
fixed effects (see below); and $\eta_{ij}$ is an error term,
statistically independent on the regressors, s.t. its mean
conditional on all regressors is equal to 1. In our exercises below,
$X$=(AREA, POP, CPI, EXC, RM), $Z$=(LL, CONT), whereas $D$=(CTG,
COMC, COML, COL, TA, COMR), see Appendix B.

The estimation of eq. \eqref{Eq:Gravity} is not easy and it
potentially entails many difficulties. Among them, we recall the
treatment of zero-valued flows
\citep{Santos2006,Linders2006,Burger2009}, non-linearity and
heteroscedasticity \citep{Santos2006}, endogeneity and omitted-term
biases \citep{Baldwin2006}. In order to simultaneously deal with all
these problems, we estimate Eq. \eqref{Eq:Gravity} using a
zero-inflated Poisson pseudo-maximum likelihood (ZIPPML) model
\citep{Burger2009} with country-fixed effects $C_{i}$
\citep{Baldwin2006} to overcome endogeneity due to omitted terms
(e.g., related to non-observable variables as landed prices of
origin goods in destination country, which may in principle be
correlated with trade-cost terms, as proxied by distance)
\footnote{Note that all variables have been original deflated, but
this does not create any difficulty here as we undertake a
cross-section study.}. To double-check our results, we also estimate
Eq. \eqref{Eq:Gravity} with alternative econometric approaches
discussed in the relevant literature \citep{Linders2006}, including
standard OLS on the log-linearized form (omitting zero-valued flows
or substituting zero-valued flows by a small constant), log-normal
pseudo-maximum likelihood estimators as in Ref. \citep{Santos2006},
and zero-inflated negative-binomial pseudo-maximum likelihood
(ZINBMML) techniques \citep{Burger2009}. Overall, we obtain the best
fit using the ZIPPML estimator, although our results are not
dramatically different under alternative estimation strategies, both
in terms of model selection and correlation of residuals, a tendency
already documented in Ref. \citep{Linders2006}.

\begin{center}
--- TABLE \ref{Tab:Gravity_Est} ABOUT HERE ---
\end{center}

Table \ref{Tab:Gravity_Est} reports the ZIPPML fit to year-2000 data, together
with usual goodness-of-fit statistics \footnote{The ZIPPML is a two-stage
procedure \citep{Wooldridge2001}. The first stage features a logit regression
estimating the probability that there is no bilateral trade at all. In the
second stage, a Poisson regression is run to fit trade flows of the non-zero
group of link (as estimated in the first stage). We use the same regressors
(Appendix B) to model both the first and second stage. Here we show only the
estimation results for the Poisson final stage for the sake of exposition (the
complete set of regression results is available on request from the Author).
All exercises have been performed using Stata 9 \citep{StataCorp}.}. The final
model has been selected by successively removing  the regressors that were not
contributing a significant impact on the overall likelihood, using a
general-to-specific procedure. Only GDP, DIST, AREA, POP, LL, CTG, COML, COL,
TA resisted this successive selection. The final gravity equation seems
well-specified, according to both Wald \citep{Wooldridge2001} and Vuong
\citep{Vuong1989} test statistics, and achieves an extremely large adjusted
$R^2$ (0.93) ---a typical performance of trade gravity equations. All signs and
magnitudes of estimated coefficients are in line with those found in the
relevant literature. Trade flows are positively affected by country GDP,
existence of trade agreements, common borders, common language and colony
relationships. They are negatively affected by distance, area and population
(net of GDP effects), as well as the probability of being landlocked. Country
remoteness, continental position, religion, exchange rates and CPI-effects do
not appear instead to significantly affect trade flows in our data.

Define $\hat{\eta}_{ij}$ as the estimated residual from eq. \eqref{Eq:Gravity},
i.e. obtained by substituting unknown parameters with estimated ones from Table
\ref{Tab:Gravity_Est}. Note that $\hat{\eta}_{ij}$ can be interpreted as the
weight of the $ij$ trade link once all structural effects related to  country
size, geographical, social, historical and political factors have been removed
from the original weight $w_{ij}$. It is then straightforward to define the
``residual'' weighted ITN by simply re-weight links using the
$\hat{\eta}_{ij}$s \footnote{See Ref. \citep{Krempel2003} for a germane
approach aimed at visualizing the properties of the ITN.}. In what follows, we
will then study the topological properties of the residual symmetric weight
matrix $E=\{e_{ij}\}$, where $e_{ij}=\hat{\eta}_{ij}$ for $i>j$,
$e_{ji}=e_{ij}$ for $i<j$, and $e_{ii}=0$ for all $i$, and compare them with
those of $W$.

\section{Topological Properties of Gravity-Equation Residual Networks}\label{Sec:Results}

\begin{figure}[h]
\begin{center}
\begin{minipage}[t]{7.5cm}
\includegraphics[width=7.5cm]{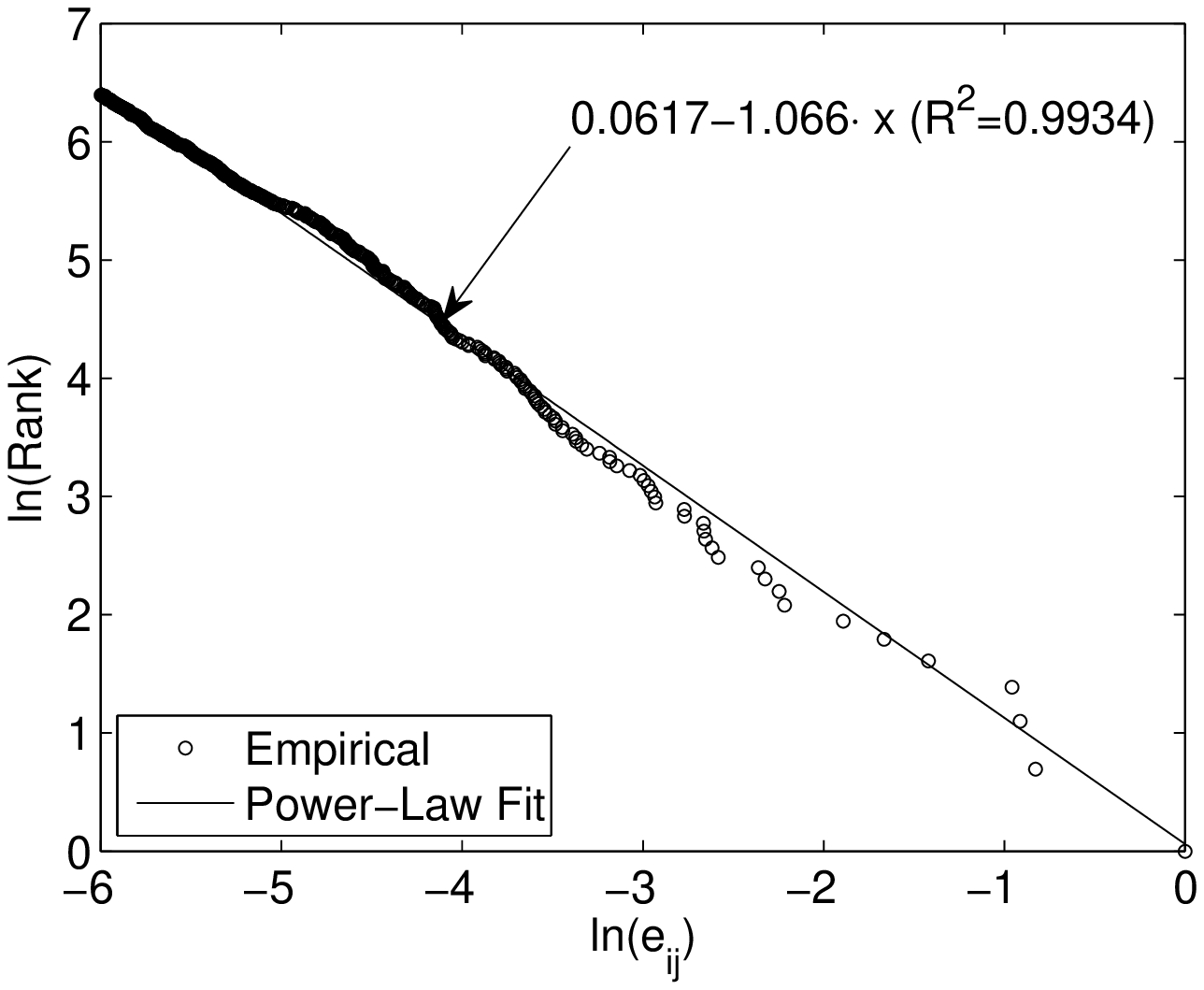}
\caption{Residual link-weight distribution. Size-rank plot with power-law fit, estimated equation and $R^2$ of the fit. Note: Estimated coefficients refer to original values and not to the rank transformation.}
\label{Fig:e_srp}
\end{minipage}
\begin{minipage}[t]{1cm}
\
\end{minipage}
\begin{minipage}[t]{7.5cm}
\includegraphics[width=7.5cm]{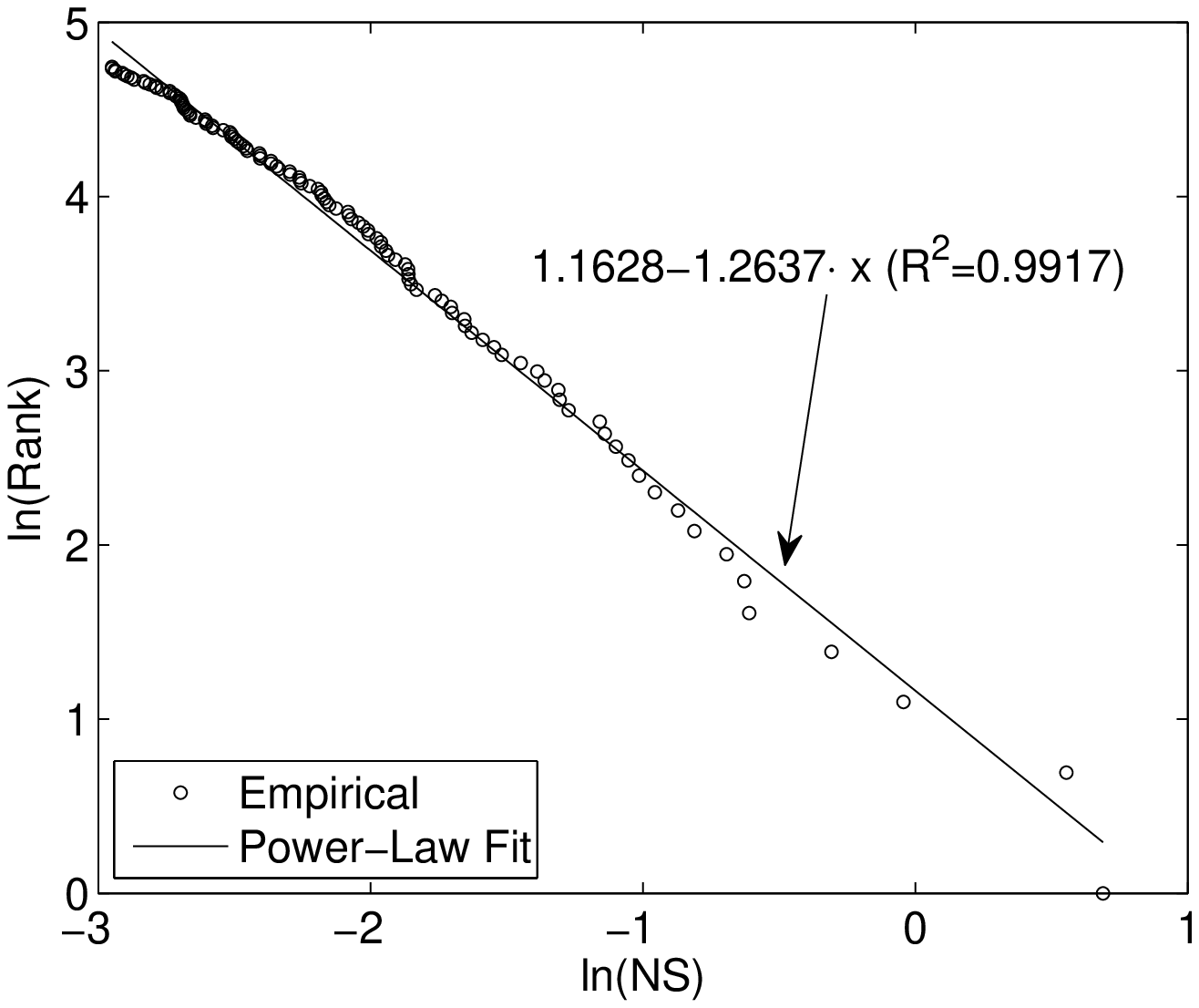}
\caption{Residual node-strength (NS) distribution. Size-rank plot with power-law fit, estimated equation and $R^2$ of the fit. Note: Estimated coefficients refer to original values and not to the rank transformation.}
\label{Fig:ns_srp}
\end{minipage}
\end{center}
\end{figure}

One of the most puzzling stylized facts emerging from the study of
the topological properties of $W$ is that the distribution of link
weights $w_{ij}$ is well approximated by a log-normal density
\citep{LiC03,Fagiolo2009pre}. Indeed, this militates against the
view that the ITN is a complex network, as log-normality can be
simply the limit outcome of uncorrelated link-weight multiplicative
growth processes (e.g., Gibrat laws; see Ref. \citep{Sutton1997}).
In other words, original link weights are markedly heterogeneous but
display exponentially-decaying upper tails, without the typical
fat-tailed behavior that is known to be the signature of complexity
\citep{Mitzenmacher2004,Rosser2008}. The present analysis shows
that, once all gravity-equation dependence is removed from the
original data, the residual ITN is characterized by power-law
distributed link weights, see the size-rank plot
\citep{Stanley_etal_1995} in Figure \ref{Fig:e_srp}. In fact, the
correlation between original and residual weights is not
statistically-significantly different from zero \ \ (-0.009, with a
p-value 0.9171). We shall go back to the implications of this
finding on modeling below. For the moment, let us stress the fact
that power-law behavior of residual link weights hints to an
inherent complex behavior of trade flows, possibly due to deep
similarities between countries, which are somewhat hidden by the
standard determinants of trade accounted for in the empirical
literature.

We now explore how the other topological properties of the residual
ITN compare to original ones. As it is customary in this literature,
we shall focus on node strength (NS), defined as the sum of all link
weights of a node; node average nearest-neighbor strength (ANNS),
i.e. the average strength of the trade partners of that node;
weighted node-clustering (WCC), measured as the relative weighted
intensity of trade triangles with that node as one of the vertices;
and random-walk betweenness centrality (RWBC), accounting for global
centrality of a node in the weighted network \footnote{See Appendix
C for formal definitions. Notice also that the ZIPPML estimator
seems to consistently predict the density of the residual ITN.
Indeed, the fraction of all possible links with positive weight is
0.63 under $W$ and 0.62 in the residual graph $E$.}.

As happens for link weights, power-law shapes characterize in the
residual ITN all node statistics (NS, ANNS, and WCC) that were
originally well-proxied by log-normal densities
\citep{Fagiolo2009pre}, see Figure \ref{Fig:ns_srp} for NS.
Interestingly, the only topological property that was power-law
shaped in the original ITN (RWBC) keeps the same shape also in the
residual network. RWBC is actually a peculiar statistics, because,
unlike the others, each of its node values somewhat reflect the
whole structure of the network. The fact that RWBC is still
power-law distributed suggests that complexity is really an
intrinsic feature of the ITN.

\begin{center}
--- TABLE \ref{Tab:CorrStructure} ABOUT HERE ---
\end{center}

Another robust set of stylized facts regarding the weighted ITN
concerns the correlation structure among node-distributions of
topological statistics, and between node topological statistics and
country per-capita GDP (as a proxy for country income). Table
\ref{Tab:CorrStructure} reports such a correlation structure for
both the original ITN ($W$) and the residual one ($E$). Coefficients
not statistically-significantly different from zero are marked in
boldface. The original ITN, where all correlation coefficients were
statistically different from zero, hinted to a trade structure where
countries that trade more intensively are also high-income ones,
they are more clustered and central, but tend to trade with
relatively-less connected partners. This configures a relatively
disassortative pattern for the ITN. Once size,
geographically-related and other determinants of trade have been
removed from the data, however, the topological properties of the
residual ITN are almost uncorrelated with their original
counterpart. The only exception is ANNS, which displays a strong and
negative correlation. This suggests that countries that in $W$
typically traded with intensively connected partners (i.e., small
and poor countries) exhibit in $E$ small ANNS values, i.e. tend to
trade with poorly-connected partners, a pattern that can be
intuitively explained by recalling that trade flows in $E$ do not
reflect any size, geographic, or colonial preferential relationship.
Note also that in $E$ countries that trade more intensively (i.e.,
high NS) still are more central and clustered, but do not display
any assortative or disassortative pattern anymore, as the NS-ANNS
correlation in $E$ is not significantly-different from zero.
Furthermore, the removal of size effects (GDP and population)
naturally destroys any positive correlation between income and trade
intensity, centrality and clustering: now high-income countries tend
to trade relatively less intensively and occupy less central
positions (and trade with relatively more connected partners).

\begin{figure}
\begin{center}
\begin{minipage}[t]{8.5cm}
\includegraphics[width=8.5cm]{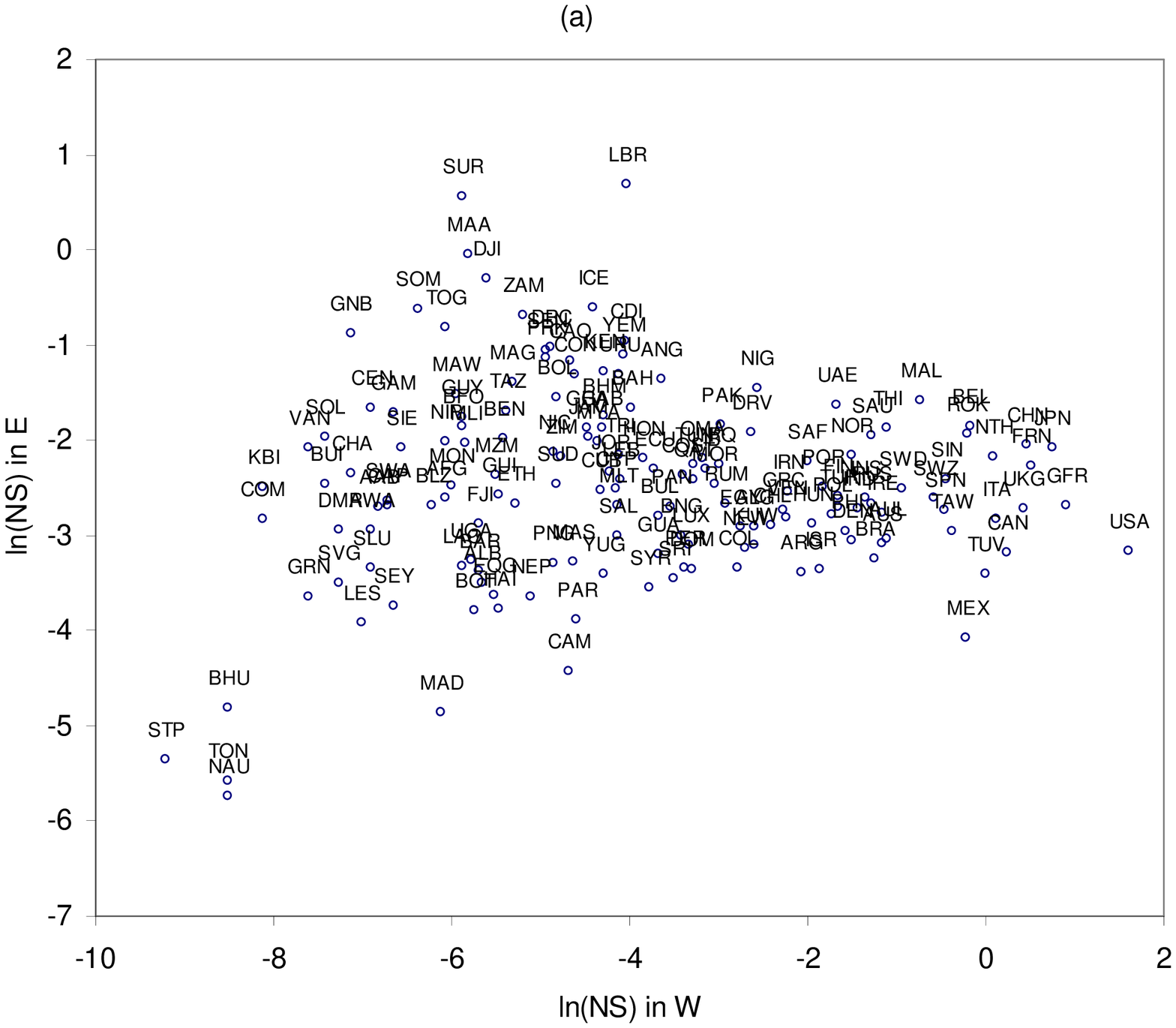}
\end{minipage}
\begin{minipage}[t]{8.5cm}
\includegraphics[width=8.5cm]{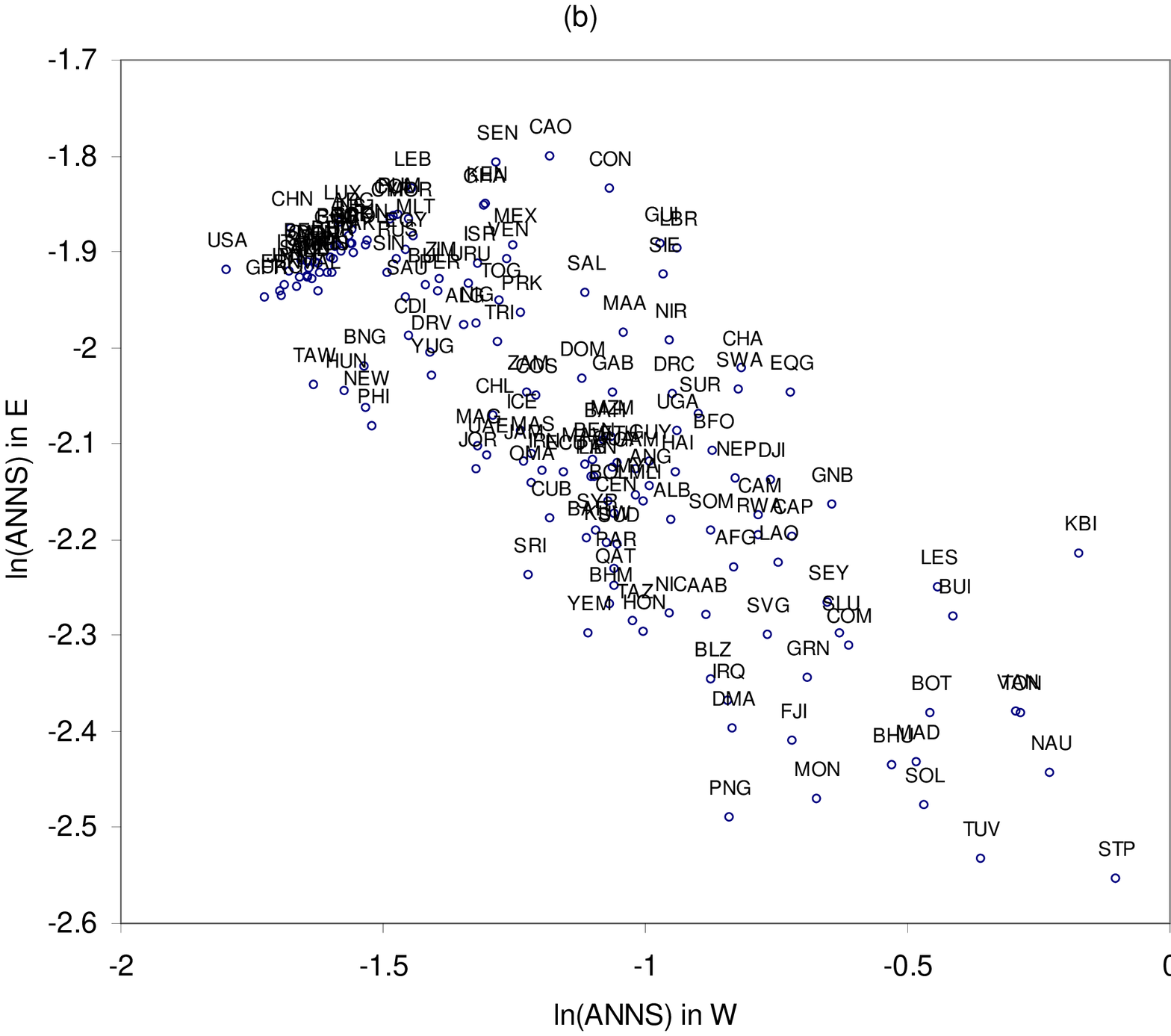}
\end{minipage}
\begin{minipage}[t]{8.5cm}
\includegraphics[width=8.5cm]{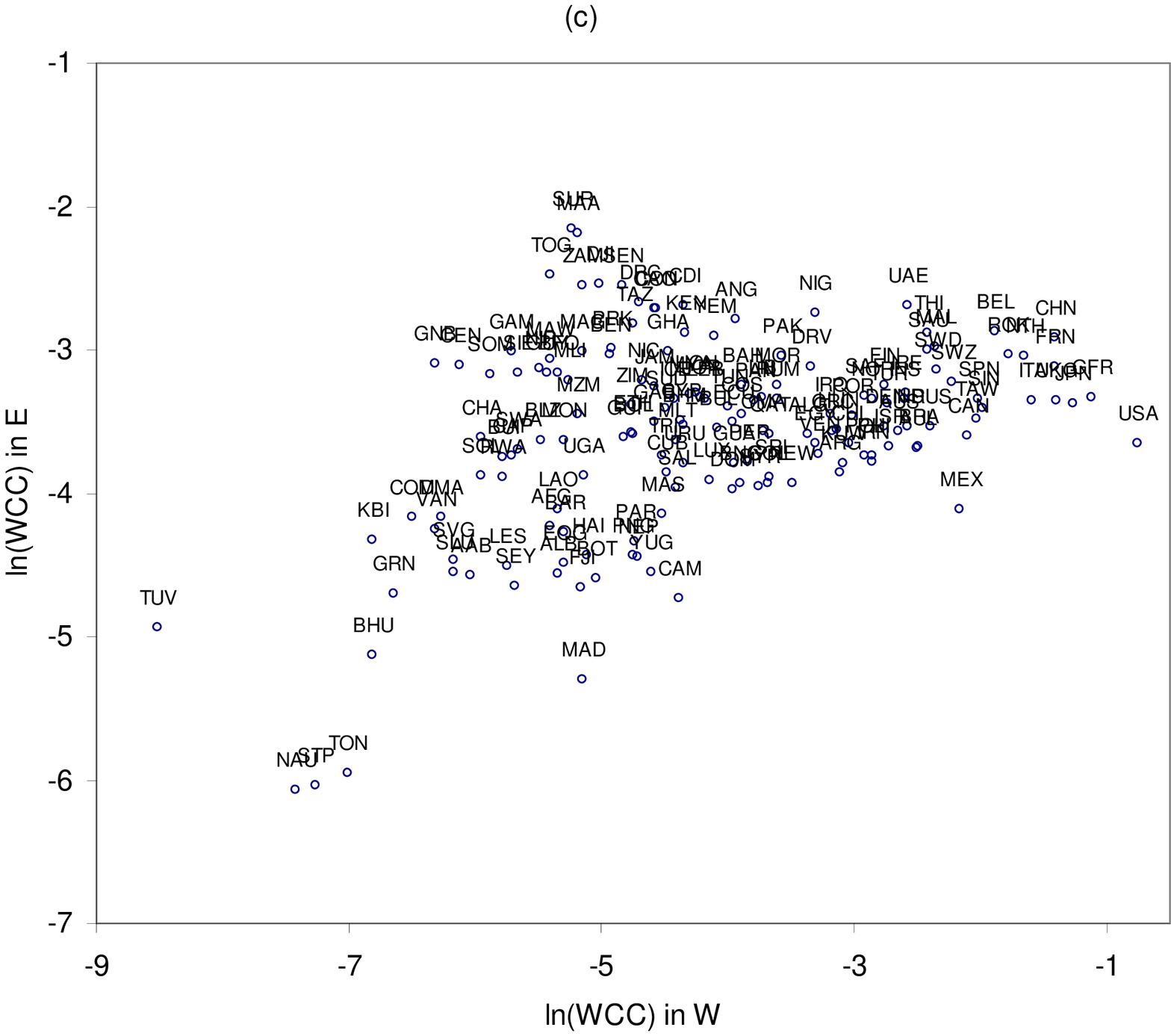}
\end{minipage}
\begin{minipage}[t]{8.5cm}
\includegraphics[width=8.5cm]{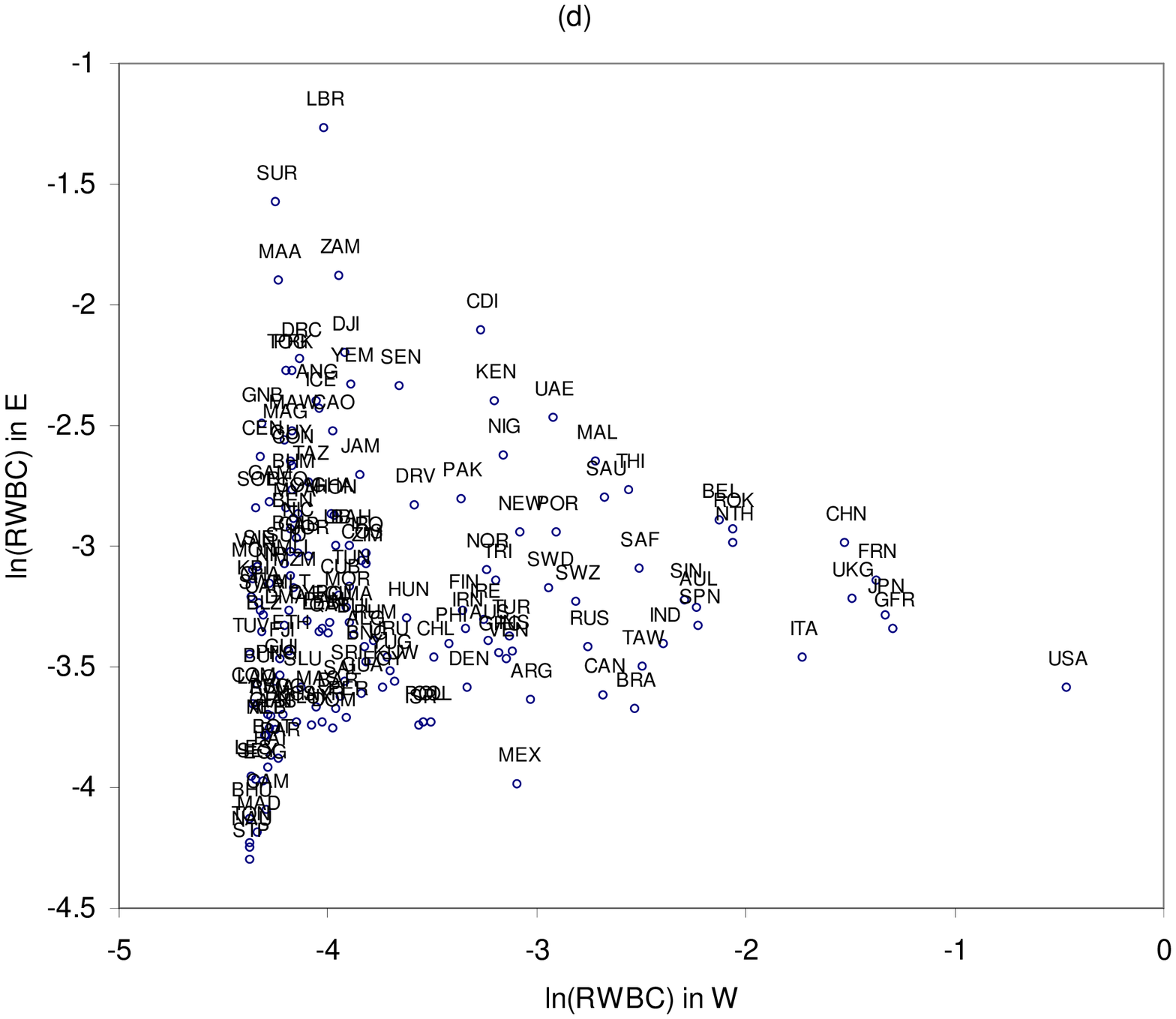}
\end{minipage}
\end{center}
\caption{Topological properties of the ITN in original vs. residual
network. (a) NS; (b) ANNS; (c) WCC; (d) RWBC. Year: 2000.}
\label{Fig:Topo_Scatters}
\end{figure}

This result is confirmed by looking at the Spearman's
rank-correlation coefficient between country-statistic rankings. If
one correlates, e.g., the rank of countries under $W$ and $E$
according to node statistics, it emerges that rankings made with
respect to NS, WCC and RWBC are only weakly (positively) correlated,
whereas the correlation for ANNS rankings is equal to -0.7676. This
suggests that if a country was scoring high in terms of intensity of
trade (NS), clustering (WCC) and centrality (RWBC) in the original
ITN, it is not likely to keep its top position in the residual
network, and with high probability will appear at the bottom of the
list in the ANNS ranking, see Figure \ref{Fig:Topo_Scatters}. For
example, whereas the US, Germany and Japan used to occupy top
positions in the rankings of trade intensity, clustering and
centrality, in the residual network such positions are now filled by
relatively small but very dynamic countries like Iceland, Korea,
Belgium, as well as middle-east oil-related countries (United-Arab
Emirates) and many African ones. Furthermore, the ANNS ranking,
which was topped in $W$ by micro-economies (e.g., Pacific islanders)
now features in the first positions relatively large and active
countries (Cameroon, Senegal, Kenya, Rumania, Colombia, Morocco,
South Africa, Mexico), as well as China and Indonesia.

All in all, the foregoing analysis shows that only few of the stylized facts
found for the weighted ITN apply to its residual version, and that a great
extent of ITN topological features can be explained by the control variables
employed in the gravity-equation regression. Our results also indicate that,
removing such structure from the data and interpreting residuals as proxies for
the underlying trade similarities between countries, interesting patterns
emerge about the inherent complex structure of the ITN and how countries are
interconnected and play their roles in the network.

Additional insights on these relationships can be gathered from a
simple weighted-graph visualization of the original vs. residual
ITN. Figures \ref{Fig:ITN_map1}-\ref{Fig:ITN_map2} depict a partial
view (largest 1\% links in terms of weights) of the original and
residual ITN in year 2000 as an undirected graph where the thickness
of a link is proportional to its weight, node size is proportional
to country's GDP, and node shapes represent the continent which the
country belongs to. It is easy to see that, as discussed in the
previous sections, large countries tend to trade more intensively
with each other, with contiguous/close countries or with partners
that are relevant in terms of history and trade agreements. The
original ITN is also characterized by a core-periphery structure
where large influential countries (USA, Japan, China, Germany,
France, Italy, etc.) play the role of hubs, attached to many other
countries by relatively weaker trade links. Geographical clustering
is also evident, as same-continent countries tend to be tightly
interconnected.

Simple inspection of the residual weighted ITN graph (links with
largest 1\% of weight in $E$) hints instead to a much more
interconnected pattern, characterized by much more spread links,
without large-sized hubs, nor excessive geographical clustering. All
large-GDP countries disappear from the graph (e.g., USA, Germany,
Italy, France). Small and medium-sized African, South-American and
Asian countries are most represented, possibly because of their
endowments in tradable natural resources and/or over-specialization
of their export profile.

Some countries, like Liberia, Surinam, Mauritania and Djibuti, play an
unexpectedly prominent role in the network. These are countries with relatively
low per-capita GDPs, which in the original network were holding many trade
relationships (between 52 and 88) and scored around the median in all
node-statistic rankings (except centrality). The majority of their trade
relationships were weak but occurred outside trade agreements. Unlike Pacific
islanders, such flows did not necessarily concern large-sized partners, were
not related to colony relationships and not necessarily geographically close
(they are not landlocked). Therefore, it is not a surprise that their residual
link weights end up being relatively large for many existing relationships,
thus promoting such countries to more important positions in the network.

Finally, to get a more precise feeling of the role of large-sized
countries in the residual network, we plot the complete minimal
spanning trees (MSTs) associated to $W$ and $E$ \footnote{See Ref.
\citep{GowerRoss1969}. To build the MST of a weighted graph with
link-weights in the unit interval we have employed the following
procedure. First, assume that information from link weights can be
treated the same way as correlation coefficients in Ref.
\citep{Mantegna1999}, i.e. that higher (symmetric) link weights
signal a higher trade similarity between countries. Second,
transform weight (i.e., similarity) matrices entries $w_{ij}$ and
$e_{ij}$ into, respectively, $w^R_{ij}=\sqrt{2(1-w_{ij})}$ and
$e^R_{ij}=\sqrt{2(1-e_{ij})}$, and let $w^R_{ii}=e^R_{ii}=0$. As
explained in Ref. \citep{Mantegna1999}, this is an appropriate
metric distance. Third, we have computed the MST associated to $W^R$
and $E^R$, and re-scaled the link weights returned by the Kruskal's
algorithm \citep[][Ch. 23.2]{Cormen01} by their maximum values, in
line with what we have done for $W$ and $E$. Finally, we have
weighted any resulting link by $1-w^R_{ij}$ and $1-e^R_{ij}$, to get
the right proportionality between weights and similarity.}. This
graphical analysis may complement the former because, by simplifying
the structure of the network, allows one to simultaneously visualize
all existing nodes and their relevant interactions, a thing that was
simply impossible to do with $W$ and $E$, given the extremely large
density of the two graphs. The plot of $W$-MST (Figure
\ref{Fig:ITN_mst1}) magnifies the hub-role of the largest countries,
especially the US, and better accounts for the position of other
peripheral countries in the ITN. Conversely, the plot of $E$-MST
(Figure \ref{Fig:ITN_mst2}) hints to a relatively more central role
of very small countries like Liberia and Surinam (see above);
downplays the importance of hubs and geography; and relegates large
countries at the periphery. For instance, US and China now link to
small countries in the ITN, whereas India keeps having a marginal
position.

\section{Conclusions}\label{Sec:Conclusions}

In this paper we have begun to investigate the determinants of the
statistical features of the weighted international-trade network. We
have compared the topological properties of the ITN, as originally
explored in a series of papers
\citep{LiC03,Bhatta2007a,Bhatta2007b,Garla2007,Fagiolo2008physa,Fagiolo2009pre},
to those of the residual weighted symmetric ITN. The latter has been
obtained after fitting original bilateral-trade flows via a standard
gravity equation including as regressors country GDP, population and
area; geographical distance between countries; and a series of link-
or country-specific variables accounting for other factors related
to geography (common border, landlocking, continent), the economy
(trade agreements, exchange rates, consumer price indices), and
social/ political/ historical effects (common language, religion,
and currency).

Our findings indicate that the residual ITN is characterized by
power-law shaped distributions of link weights and node statistics
(e.g., strength, clustering, random-walk betweenness centrality).
Hence, the underlying architecture of the weighted ITN seems to
exhibit signatures of complexity, unlike its original counterpart.
We have also found that the correlation structure among node
statistics, and between node statistics and country per-capita GDP,
changes substantially when comparing the original and residual
weighted networks. Whereas the original ITN displays a structure
with a few large-sized country hubs and a relatively strong
connectivity among nearby countries (either geographically,
socially, politically or economically), the residual ITN is
organized around many relatively small-sized but trade-oriented
countries that, independently of their geographical position, either
play the role of local hubs or attract large and rich countries in
complex trade-interaction patterns.

Our results have some implications for modeling purposes. Indeed,
the strong gravity-like dependence that, in line with existing
literature, can be detected in ITN data, as well as the impact of
removing such structure on ITN topological properties, suggests that
any model of network formation and link-weight evolution aiming at
replicating ITN topological properties should consider country size,
geography, etc. among its building blocks. This may occur, for
instance, by devising dynamic models of network formation and weight
evolution where node size and position, contiguity and agreements,
matter in the decision of the nodes to form/delete a link, or to
change its weight. At the same time, models aiming at explaining and
replicating ITN stylized facts should allow empirical calibration of
their node- and link-related characteristics (see, Refs.
\citep{Garla2004,Garla2005,Garla2007,Bhatta2007a} for preliminary
exercises in this direction). This may be relevant for predictive
and policy exercises. Furthermore, our findings call for models that
are simultaneously able to replicate the set of stylized facts of
both the original and the residual ITN (no matter the determinants
accounted for in the gravity equation). For example, it would be
important to have models of the ITN that recover log-normality of
the (equilibrium) link-weight distribution when size and geography
are not accounted for, and power-law shapes after having controlled
for them.

To conclude, it may be worthwhile to notice that the simple and
preliminary exercises described in this article may be extended in
at least two directions. First, one may take a specific-to-general
approach to gravity-equation modeling, and adding an increasing
number of factors in its specification. For example, one may begin
with size only, and gradually introduce geography, trade agreements,
etc.. At each step a residual ITN can be defined and its properties
can be accordingly compared to those of the original ITN. This
analysis may end up in a minimal ``sufficient'' set of determinants
able to, e.g., account for a complex ITN structure, or changes in
node-statistics correlation patterns. Finally, one might play with
expanded versions of the gravity equation that consider, e.g.,
country-specific natural endowments, industrial profile and trade
specialization. This might shed some light on the peculiar structure
of the residual network and explain trade patterns that otherwise
could remain obscure.


\newpage

\begin{appendix}

\section{List of Countries in the Balanced Panel (1981-2000).}

{\squeezetable \begin{table}[h] \centering \label{Tab:countries}
\begin{tabular}{lllllllll}
\hline\noalign{\smallskip}
ID &       Acro &       Name &         ID &       Acro &       Name &         ID &       Acro &       Name \\
\noalign{\smallskip}\hline\noalign{\smallskip}
2 &         USA  &  United States  &        355 &         BUL  &    Bulgaria  &        600 &         MOR  &     Morocco  \\
20 &         CAN  &      Canada  &        360 &         RUM  &     Rumania  &        615 &         ALG  &     Algeria  \\
31 &         BHM  &     Bahamas  &        365 &         RUS  &      Russia  &        616 &         TUN  &     Tunisia  \\
40 &         CUB  &        Cuba  &        375 &         FIN  &     Finland  &        620 &         LIB  &       Libya  \\
41 &         HAI  &       Haiti  &        380 &         SWD  &      Sweden  &        625 &         SUD  &       Sudan  \\
42 &         DOM  &  Dominican Rep.  &        385 &         NOR  &      Norway  &        630 &         IRN  &        Iran  \\
51 &         JAM  &     Jamaica  &        390 &         DEN  &     Denmark  &        640 &         TUR  &      Turkey  \\
52 &         TRI  &  Trinidad/Tobago  &        395 &         ICE  &     Iceland  &        645 &         IRQ  &        Iraq  \\
53 &         BAR  &    Barbados  &        402 &         CAP  &  Cape Verde  &        651 &         EGY  &       Egypt  \\
54 &         DMA  &    Dominica  &        403 &         STP  &    Sao Tome  &        652 &         SYR  &       Syria  \\
55 &         GRN  &     Grenada  &        404 &         GNB  &  Guinea-Bissau  &        660 &         LEB  &     Lebanon  \\
56 &         SLU  &  Saint Lucia  &        411 &         EQG  &  Eq. Guinea  &        663 &         JOR  &      Jordan  \\
57 &         SVG  &  St. Vincent  &        420 &         GAM  &      Gambia  &        666 &         ISR  &      Israel  \\
58 &         AAB  &     Antigua  &        432 &         MLI  &        Mali  &        670 &         SAU  &  Saudi Arabia  \\
70 &         MEX  &      Mexico  &        433 &         SEN  &     Senegal  &        678 &         YEM  &       Yemen  \\
80 &         BLZ  &      Belize  &        434 &         BEN  &       Benin  &        690 &         KUW  &      Kuwait  \\
90 &         GUA  &   Guatemala  &        435 &         MAA  &  Mauritania  &        692 &         BAH  &     Bahrain  \\
91 &         HON  &    Honduras  &        436 &         NIR  &       Niger  &        694 &         QAT  &       Qatar  \\
92 &         SAL  &  El Salvador  &        437 &         CDI  &  Cote D'Ivoire  &        696 &         UAE  &  Arab Emirates  \\
93 &         NIC  &   Nicaragua  &        438 &         GUI  &      Guinea  &        698 &         OMA  &        Oman  \\
94 &         COS  &  Costa Rica  &        439 &         BFO  &  Burkina Faso  &        700 &         AFG  &  Afghanistan  \\
95 &         PAN  &      Panama  &        450 &         LBR  &     Liberia  &        710 &         CHN  &       China  \\
100 &         COL  &    Colombia  &        451 &         SIE  &  Sierra Leone  &        712 &         MON  &    Mongolia  \\
101 &         VEN  &   Venezuela  &        452 &         GHA  &       Ghana  &        713 &         TAW  &      Taiwan  \\
110 &         GUY  &      Guyana  &        461 &         TOG  &        Togo  &        731 &         PRK  &  North Korea  \\
115 &         SUR  &     Surinam  &        471 &         CAO  &    Cameroon  &        732 &         ROK  &  South Korea  \\
130 &         ECU  &     Ecuador  &        475 &         NIG  &     Nigeria  &        740 &         JPN  &       Japan  \\
135 &         PER  &        Peru  &        481 &         GAB  &       Gabon  &        750 &         IND  &       India  \\
140 &         BRA  &      Brazil  &        482 &         CEN  &Centr African Rep.  &        760 &         BHU  &      Bhutan  \\
145 &         BOL  &     Bolivia  &        483 &         CHA  &        Chad  &        770 &         PAK  &    Pakistan  \\
150 &         PAR  &    Paraguay  &        484 &         CON  &       Congo  &        771 &         BNG  &  Bangladesh  \\
155 &         CHL  &       Chile  &        490 &         DRC  &  Congo (Zaire)  &        775 &         MYA  &     Myanmar  \\
160 &         ARG  &   Argentina  &        500 &         UGA  &      Uganda  &        780 &         SRI  &   Sri Lanka  \\
165 &         URU  &     Uruguay  &        501 &         KEN  &       Kenya  &        781 &         MAD  &    Maldives  \\
200 &         UKG  &  United Kingdom  &        510 &         TAZ  &    Tanzania  &        790 &         NEP  &       Nepal  \\
205 &         IRE  &     Ireland  &        516 &         BUI  &     Burundi  &        800 &         THI  &    Thailand  \\
210 &         NTH  &  Netherlands  &        517 &         RWA  &      Rwanda  &        811 &         CAM  &    Cambodia  \\
211 &         BEL  &     Belgium  &        520 &         SOM  &     Somalia  &        812 &         LAO  &        Laos  \\
212 &         LUX  &  Luxembourg  &        522 &         DJI  &    Djibouti  &        816 &         DRV  &     Vietnam  \\
220 &         FRN  &      France  &        530 &         ETH  &    Ethiopia  &        820 &         MAL  &    Malaysia  \\
225 &         SWZ  &  Switzerland  &        540 &         ANG  &      Angola  &        830 &         SIN  &   Singapore  \\
230 &         SPN  &       Spain  &        541 &         MZM  &  Mozambique  &        840 &         PHI  &  Philippines  \\
235 &         POR  &    Portugal  &        551 &         ZAM  &      Zambia  &        850 &         INS  &   Indonesia  \\
260 &         GFR  &     Germany  &        552 &         ZIM  &    Zimbabwe  &        900 &         AUL  &   Australia  \\
290 &         POL  &      Poland  &        553 &         MAW  &      Malawi  &        910 &         PNG  &       Papua  \\
305 &         AUS  &     Austria  &        560 &         SAF  &  South Africa  &        920 &         NEW  &  New Zealand  \\
310 &         HUN  &     Hungary  &        570 &         LES  &     Lesotho  &        935 &         VAN  &     Vanuatu  \\
325 &         ITA  &       Italy  &        571 &         BOT  &    Botswana  &        940 &         SOL  &   Solomon's  \\
338 &         MLT  &       Malta  &        572 &         SWA  &   Swaziland  &        950 &         FJI  &        Fiji  \\
339 &         ALB  &     Albania  &        580 &         MAG  &  Madagascar  &        970 &         KBI  &    Kiribati  \\
345 &         YUG  &  Yugoslavia  &        581 &         COM  &     Comoros  &        971 &         NAU  &       Nauru  \\
350 &         GRC  &      Greece  &        590 &         MAS  &   Mauritius  &        972 &         TON  &       Tonga  \\
352 &         CYP  &      Cyprus  &        591 &         SEY  &  Seychelles  &        973 &         TUV  &      Tuvalu  \\
\noalign{\smallskip}\hline
\end{tabular}
\end{table}}

\newpage

\section{List of link- or country-related additional variables employed in gravity-equation exercises.}

{\squeezetable \begin{table}[h] \label{Tab:gravity_vars} \centering
\begin{tabular}{p{1cm}p{1.5cm}p{5cm}p{5cm}}

     Label & Related to & Explanation &     Source \\
\hline
       GDP &    Country & Gross-domestic product & Gleditsch (2002) \\
 & & \\
      AREA &    Country & Country area in Km$^2$ & CEPII (http://www.cepii.fr/) \\
& & \\
       POP &    Country & Country population & Gleditsch (2002) \\
& & \\
        LL &    Country & Dummy variable equal to 1 for landlocked countries & CEPII (http://www.cepii.fr/) \\
& & \\
      CONT &    Country & Dummy variables recording the continent to which the country belongs & CEPII (http://www.cepii.fr/) \\
& & \\
        RM &    Country & Country remoteness index defined as the weighted average of the distances of a country to all other countries, with weights equal to the country share of world GDP, see \citep{Bhavnani2002}. & Our computations on CEPII and Gleditsch (200) data \\
& & \\
       CPI &    Country & Consumer price index & International Monetary Fund (www.imf.org) \\
& & \\
      DIST &       Link & Geodesic geographical distance between two countries, calculated with the great circle formula & CEPII (http://www.cepii.fr/) \\
& & \\
       CTG &       Link & Contiguity dummy equal to 1 if two countries share a common border & CEPII (http://www.cepii.fr/) \\
& & \\
      COMC &       Link & Dummy equal to 1 if two countries use the same currency & Our computation on Andrew Rose dataset (http://faculty.haas.berkeley.edu/arose), see also \citep{GlickRose2001}. \\
& & \\
      COML &       Link & Dummy equal to 1 if the official language (or mother tongue or second language) of the two countries is the same & CEPII (http://www.cepii.fr/) \\
& & \\
       COL &       Link & Dummy equal to 1 if the two countries share a substantial colonizer-colonized relationship & CEPII (http://www.cepii.fr/) \\
& & \\
        TA &       Link & Dummy variable equal to 1 for countries involved in regional, bilateral or preferential trade agreements in year 1995 still in place in year 2000 & WTO (http://www.wto.org/) \\
& & \\
       EXC &       Link & Nominal exchange rates & International Monetary Fund (www.imf.org) \\
& & \\
      COMR &       Link & Dummy equal to 1 if the two countries share a common religion & Our computations on Andrew Rose dataset (http://faculty.haas.berkeley.edu/arose), see also \citep{RoseSpiegel2002}. \\
\hline
\end{tabular}
\end{table}}

\newpage

\section{Network Statistics}

Given a $N\times N$ symmetric weight matrix $W=\{w_{ij}\}$, with $0\leq w_{ij}\leq 1$, define the associated symmetric adjacency matrix as $A=\{a_{ij}\}$, where $a_{ij}=1$ iff $w_{ij}>0$ and zero otherwise. In the paper, we make use of the following statistics:

\begin{itemize}
    \item \textit{Node degree} \citep{AlbertBarabasi2002}, defined as
  $ND_i=A_{(i)}\textbf{1}$, where $A_{(i)}$ is the $i$-th row of $A$
  and $\textbf{1}$ is a unary vector. ND is a measure of binary
  connectivity, as it counts the number of trade partners of any
  given node. Although we mainly focus here on a weighted-network
  approach, we study ND because of its natural interpretation in
  terms of number of trade partnerships and bilateral trade agreements.

\item \textit{Node strength} \citep{DeMontis2005}, defined as $NS_i=W_{(i)}\textbf{1}$, where $W_{(i)}$ is the $i$-th row of $W$. NS is a measure of weighted connectivity, as it gives us an idea of how intense existing trade relationships of country $i$ are.

  \item \textit{Node average nearest-neighbor strength}
      \citep{DeMontis2005}, that is $ANNS_i=(A_{(i)} W\textbf{1})$
      $/(A_{(i)}\textbf{1})$. ANNS measures how intense are trade
      relationships maintained by the partners of a given node. Therefore,
      the correlation between ANNS and NS is a measure of network
      assortativity (if positive) or disassortativity (if negative).

  \item \textit{Weighted clustering coefficient} \citep{Saramaki2006,Fagiolo2007pre}, defined as
        \begin{equation*}
        {WCC_i=\frac{({W}^{\left[\frac{1}{3}\right]})_{ii}^{3}}{ND_i(ND_i-1)}.}\end{equation*}
        Here $Z_{ii}^3$ is the $i$-th entry on the main diagonal of $Z\cdot Z\cdot Z$ and
        $Z^{\left[m\right]}$ stands for the matrix obtained from $Z$ after raising each
  entry to $m$. WCC measures how much clustered a node $i$ is from a weighted perspective, i.e.
  how much intense are the linkages of trade triangles having country $i$ as a
  vertex. Replacing $W$ with $A$, one obtains the standard binary clustering coefficient
  (BCC), which counts the fraction of triangles existing in the neighborhood of any give node
  \citep{WattsStrogatz1998}.

  \item \textit{Random-walk betweenness centrality}
  \citep{Newman2005,FisherVega2006}, which is a measure of how much a given
  country is globally-central in the ITN. A node has a higher
  random-walk betweenness centrality (RWBC) the more it has a position of strategic significance
  in the overall structure of the network. In other words, RWBC is
  the extension of node betweenness centrality to weighted
  networks and measures the probability that a random signal can find its way
  through the network and reach the target node where the links to follow are
  chosen with a probability proportional to their weights.

\end{itemize}

\end{appendix}

\newpage \clearpage

{\squeezetable
\begin{sidewaystable}[htb] \centering
\begin{tabular}{rrrrrrrrrrrrrr}

           &            &            &            &            &            &            &            &            &            &            &            &            &   \% Total \\

           &  N America &  C America &  S America & Cont Europe & East Europe & Middle East &     C Asia & China \& E Asia &   N Africa &   C Africa &   S Africa &    Pacific &      Trade \\
\hline
 N America &    31.87\% &    12.35\% &     3.66\% &    18.52\% &     0.71\% &     2.68\% &     0.98\% &    26.57\% &     0.44\% &     0.68\% &     0.49\% &     1.05\% &    20.49\% \\

 C America &    74.19\% &     3.96\% &     3.37\% &     8.50\% &     0.44\% &     0.59\% &     0.15\% &     8.06\% &     0.15\% &     0.15\% &     0.15\% &     0.29\% &     3.41\% \\

 S America &    29.88\% &     4.58\% &    23.90\% &    22.11\% &     1.00\% &     1.99\% &     0.80\% &    12.35\% &     1.20\% &     1.00\% &     0.60\% &     0.60\% &     2.51\% \\

Cont Europe &     9.75\% &     0.75\% &     1.43\% &    67.53\% &     3.75\% &     3.17\% &     0.76\% &     9.44\% &     1.45\% &     0.72\% &     0.64\% &     0.60\% &    38.91\% \\

East Europe &     6.43\% &     0.67\% &     1.11\% &    64.75\% &    10.64\% &     3.99\% &     0.89\% &     9.76\% &     0.67\% &     0.67\% &     0.22\% &     0.22\% &     2.26\% \\

Middle East &    13.87\% &     0.50\% &     1.26\% &    31.15\% &     2.27\% &     9.08\% &     4.67\% &    32.53\% &     1.51\% &     0.63\% &     1.26\% &     1.26\% &     3.97\% \\

    C Asia &    16.53\% &     0.41\% &     1.65\% &    24.38\% &     1.65\% &    15.29\% &     4.55\% &    29.34\% &     1.24\% &     2.48\% &     0.83\% &     1.65\% &     1.21\% \\

China \& E Asia &    23.08\% &     1.17\% &     1.31\% &    15.58\% &     0.93\% &     5.47\% &     1.50\% &    46.82\% &     0.32\% &     0.57\% &     0.59\% &     2.65\% &    23.59\% \\

  N Africa &    10.06\% &     0.56\% &     3.35\% &    63.13\% &     1.68\% &     6.70\% &     1.68\% &     8.38\% &     2.79\% &     1.12\% &     0.00\% &     0.56\% &     0.90\% \\

  C Africa &    17.95\% &     0.64\% &     3.21\% &    35.90\% &     1.92\% &     3.21\% &     3.85\% &    17.31\% &     1.28\% &    12.18\% &     1.92\% &     0.64\% &     0.78\% \\

  S Africa &    15.04\% &     0.75\% &     2.26\% &    37.59\% &     0.75\% &     7.52\% &     1.50\% &    21.05\% &     0.00\% &     2.26\% &     9.77\% &     1.50\% &     0.67\% \\

   Pacific &    16.17\% &     0.75\% &     1.13\% &    17.67\% &     0.38\% &     3.76\% &     1.50\% &    46.99\% &     0.38\% &     0.38\% &     0.75\% &    10.15\% &     1.33\% \\
\hline
\end{tabular}
\caption{\label{Tab:Within_Between}Total trade within and between
macro geographical areas. Columns 1-12: Percentage of total trade of
countries in the area that is exchanged with countries of the same
or other areas. Column 13: Share of total trade of the area to world
total trade. Year: 2000.}
\end{sidewaystable}}

\newpage \clearpage

\begin{table}[ht] \centering
\begin{tabular}{ccccc}
\hline \hline
   Regressor &  Coefficient &            &    Regressor &  Coefficient    \\

    &  (Rob. SE) &            &     &  (Rob. SE)    \\

\hline

Log GDP$_i$ &   1.471*** &            &    LL$_i$ &  -0.336*** \\

           &    (0.000) &            &            &    (0.000) \\

Log GDP$_j$ &   1.338*** &            &    LL$_j$ &  -0.019*** \\

           &    (0.000) &            &            &    (0.000) \\

  Log DIST &  -0.727*** &            &        CTG &   0.553*** \\

           &    (0.000) &            &            &    (0.000) \\

Log AREA$_i$ &  -0.144*** &            &       COML &   0.242*** \\

           &    (0.000) &            &            &    (0.000) \\

Log AREA$_j$ &  -0.187*** &            &        COL &   0.007*** \\

           &    (0.000) &            &            &    (0.001) \\

Log POP$_i$ &  -0.504*** &            &         TA &   0.024*** \\

           &    (0.000) &            &            &    (0.000) \\

Log POP$_j$ &  -0.413*** &            &          - &          - \\

           &    (0.000) &            &          - &          - \\

\hline

  Constant &         NO &            & Wald $\chi^2$ &   17600000 \\

Country Dummies &        YES &            & Prob $>\chi^2$ &    0.00*** \\

   Adj. $R^2$ &       0.93 &            &    Vuong Z &       5.73 \\

Log Likelihood &    -631000 &            &     Prob $>$Z &    0.00*** \\
\hline \hline
\end{tabular}
\caption{\label{Tab:Gravity_Est}Estimated gravity equation for year
2000. Poisson regression. Note: Dependent variable = $w_{ij}$. No.
Observations = 12561. Estimates obtained using a ZIPPML procedure.
Robust standard errors in parentheses. Legend: $^\star: p<0.05$;
$^{\star \star}: p<0.01$; $^{\star \star \star}:
p<0.001$.}\end{table}

\vskip 3cm

\begin{table}[hb]
\centering \small \begin{tabular}{cc|cccc|cccc|c}

           &            &                            \multicolumn{ 4}{c}{W} &                            \multicolumn{ 4}{c}{E} &            \\

           &            &         NS &       ANNS &        WCC &       RWBC &         NS &       ANNS &        WCC &       RWBC &      pcGDP \\
\hline
\multicolumn{ 1}{c}{} &         NS &          - &    -0.3453 &     0.9484 &     0.5741 & {\bf -0.0881} &     0.2902 & {\bf 0.0331} & {\bf -0.0909} &     0.5170 \\

\multicolumn{ 1}{c}{W} &       ANNS &          - &          - &    -0.4774 &    -0.3759 & {\bf 0.0051} &    -0.7753 &    -0.2967 & {\bf -0.0887} &    -0.4590 \\

\multicolumn{ 1}{c}{} &        WCC &          - &          - &          - &     0.5437 & {\bf -0.1133} &     0.3985 & {\bf 0.0698} & {\bf -0.1011} &     0.5968 \\

\multicolumn{ 1}{c}{} &       RWBC &          - &          - &          - &          - & {\bf -0.0797} &     0.3178 & {\bf 0.0554} & {\bf -0.0673} &     0.4975 \\

\hline

\multicolumn{ 1}{c}{} &         NS &          - &          - &          - &          - &          - & {\bf 0.1155} &     0.8363 &     0.5202 &    -0.1678 \\

\multicolumn{ 1}{c}{E} &       ANNS &          - &          - &          - &          - &          - &          - &     0.3834 &     0.1574 &     0.3312 \\

\multicolumn{ 1}{c}{} &        WCC &          - &          - &          - &          - &          - &          - &          - &     0.5193 & {\bf -0.0961} \\

\multicolumn{ 1}{c}{} &       RWBC &          - &          - &          - &          - &          - &          - &          - &          - &    -0.1908 \\
\hline
\end{tabular}
\caption{\label{Tab:CorrStructure}Correlation structure between
topological properties, and country per-capita GDP (pcGDP). Note:
W=original weighted ITN; E=residual weighted ITN; in boldface
not-statistically significant correlation coefficients.}\end{table}

\newpage \clearpage

\begin{sidewaysfigure}[h]
\begin{center}
\includegraphics[width=20cm]{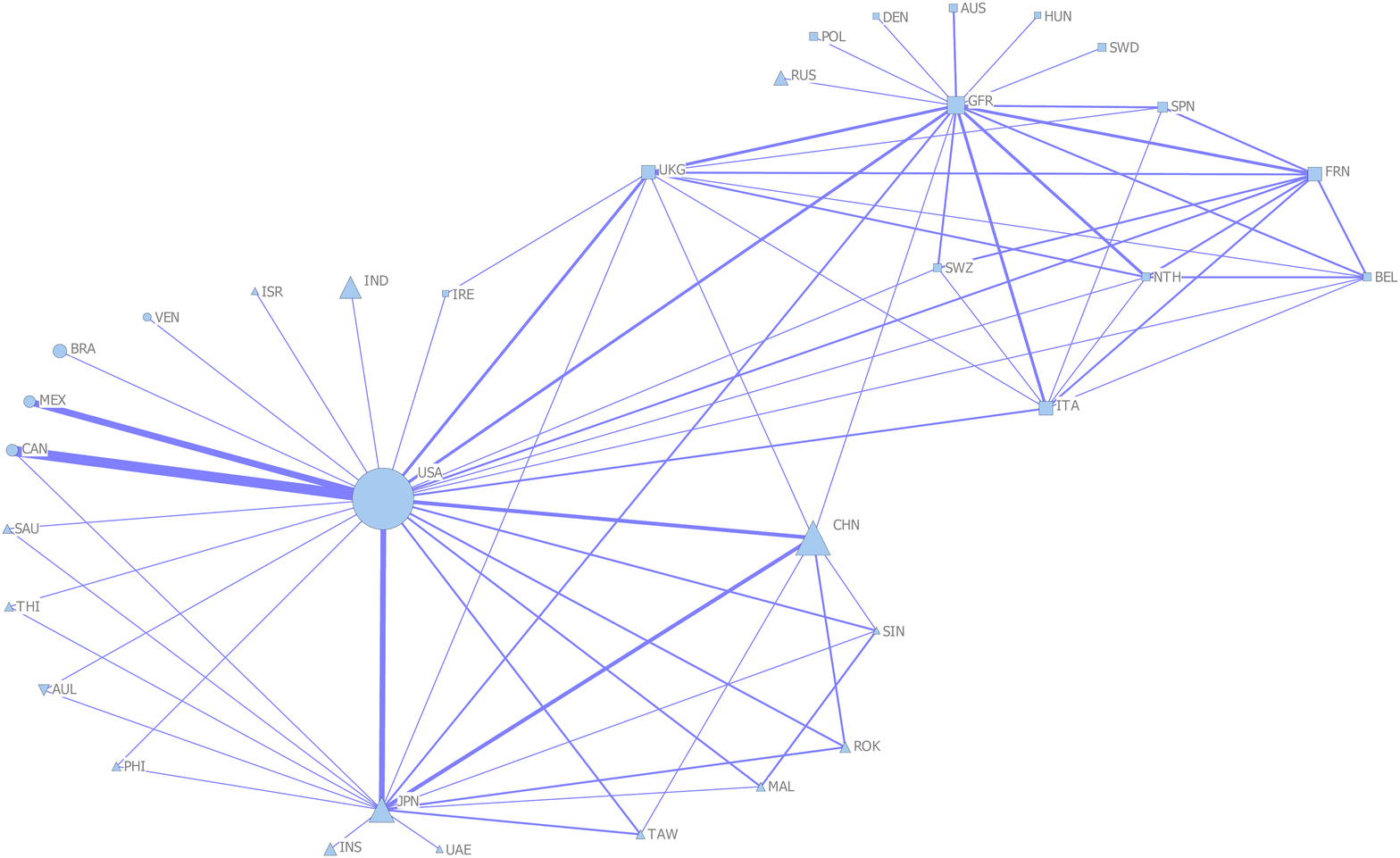}
\caption{\label{Fig:ITN_map1} A partial visualization of the
original weighted ITN ($W$). Thickness of links is proportional to
their weight. Only the largest 1\% of links are shown. Node sizes
are proportional to country's GDP. Node shapes represent the
continent which the country belongs to (Circles: America; Empty
Squares: Europe; Upright Triangles: Asia; Crossed Squares: Africa;
Reversed Triangles: Pacific).}
\end{center}
\end{sidewaysfigure}

\newpage \clearpage

\begin{sidewaysfigure}[h]
\begin{center}
\includegraphics[width=20cm]{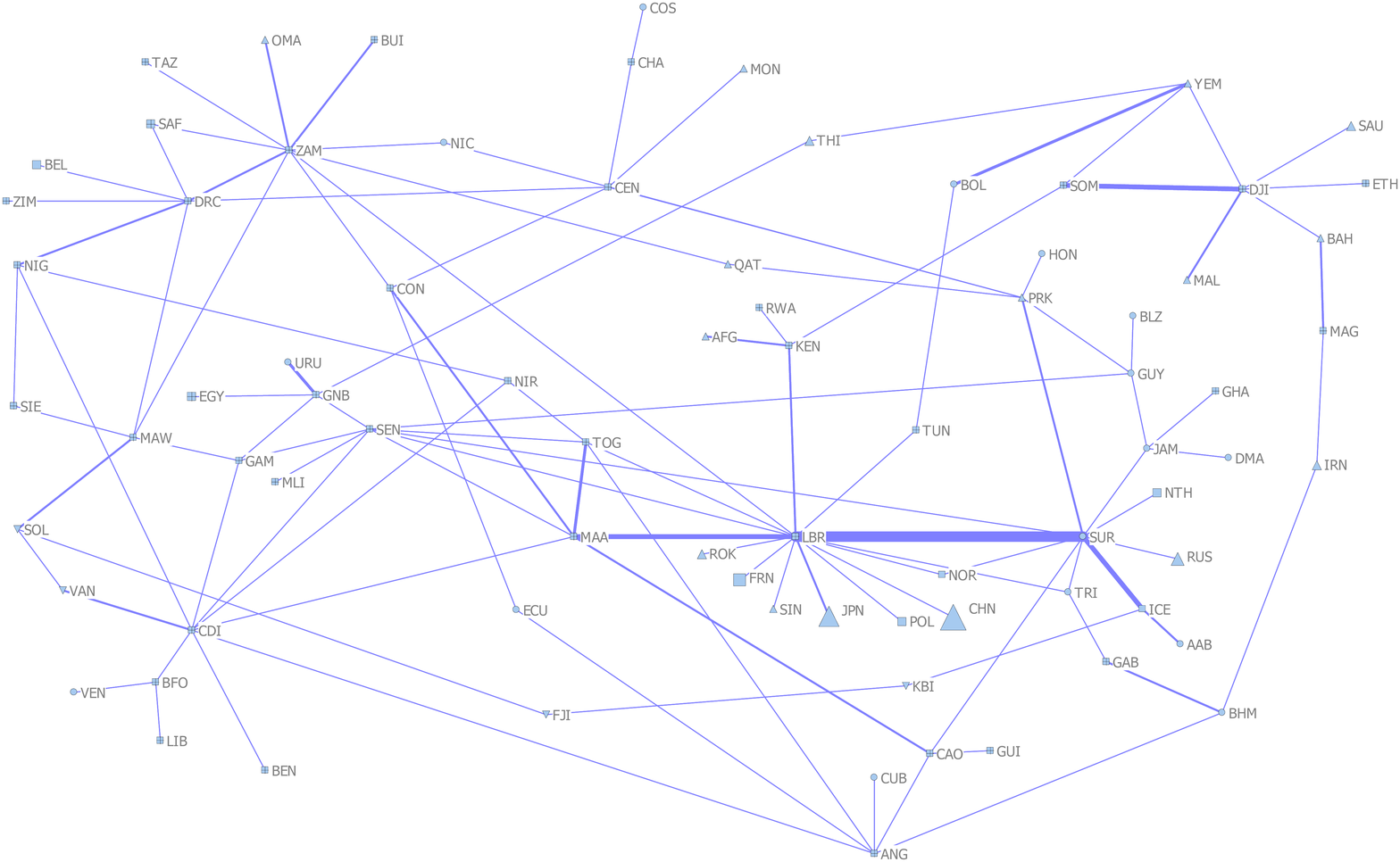}
\caption{\label{Fig:ITN_map2}A partial visualization of the residual
weighted ITN ($E$). Thickness of links is proportional to their
weight. Only the largest 1\% of links are shown. Node sizes are
proportional to country's GDP. Node shapes represent the continent
which the country belongs to (Circles: America; Empty Squares:
Europe; Upright Triangles: Asia; Crossed Squares: Africa; Reversed
Triangles: Pacific).}
\end{center}
\end{sidewaysfigure}

\newpage \clearpage

\begin{sidewaysfigure}[h]
\begin{center}
\includegraphics[width=20cm]{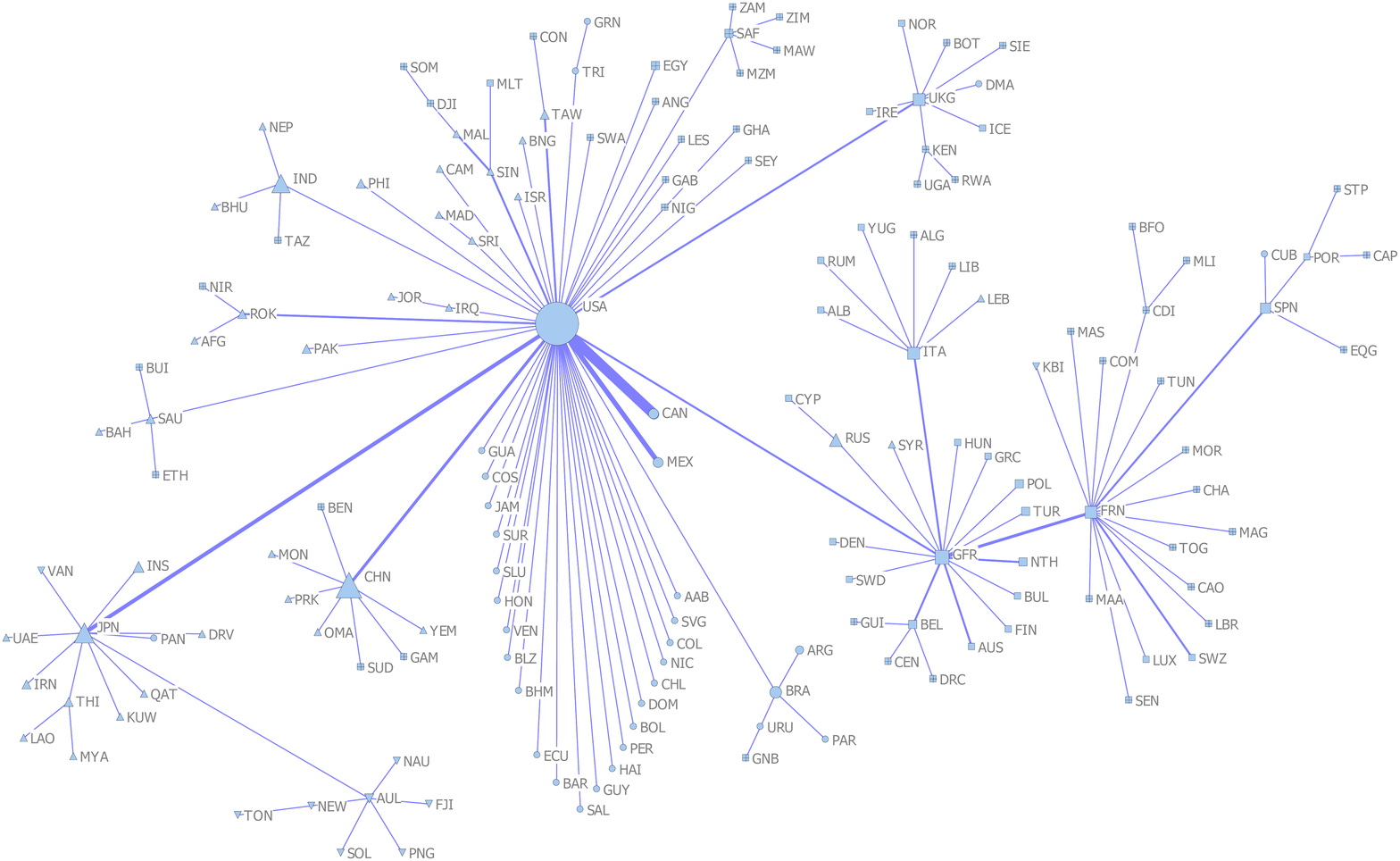}
\caption{\label{Fig:ITN_mst1}Complete minimal spanning tree (MST) of
the original weighted ITN ($W$). Thickness of a link is proportional
to one minus the weight associated to that link in the MST returned
by the Kruskal's algorithm \citep[][Ch. 23.2]{Cormen01}. Node sizes
are proportional to country's GDP. Node shapes represent the
continent which the country belongs to (Circles: America; Empty
Squares: Europe; Upright Triangles: Asia; Crossed Squares: Africa;
Reversed Triangles: Pacific).}
\end{center}
\end{sidewaysfigure}

\newpage \clearpage

\begin{sidewaysfigure}[h]
\begin{center}
\includegraphics[trim = 13 0 0 0,clip,width=20cm]{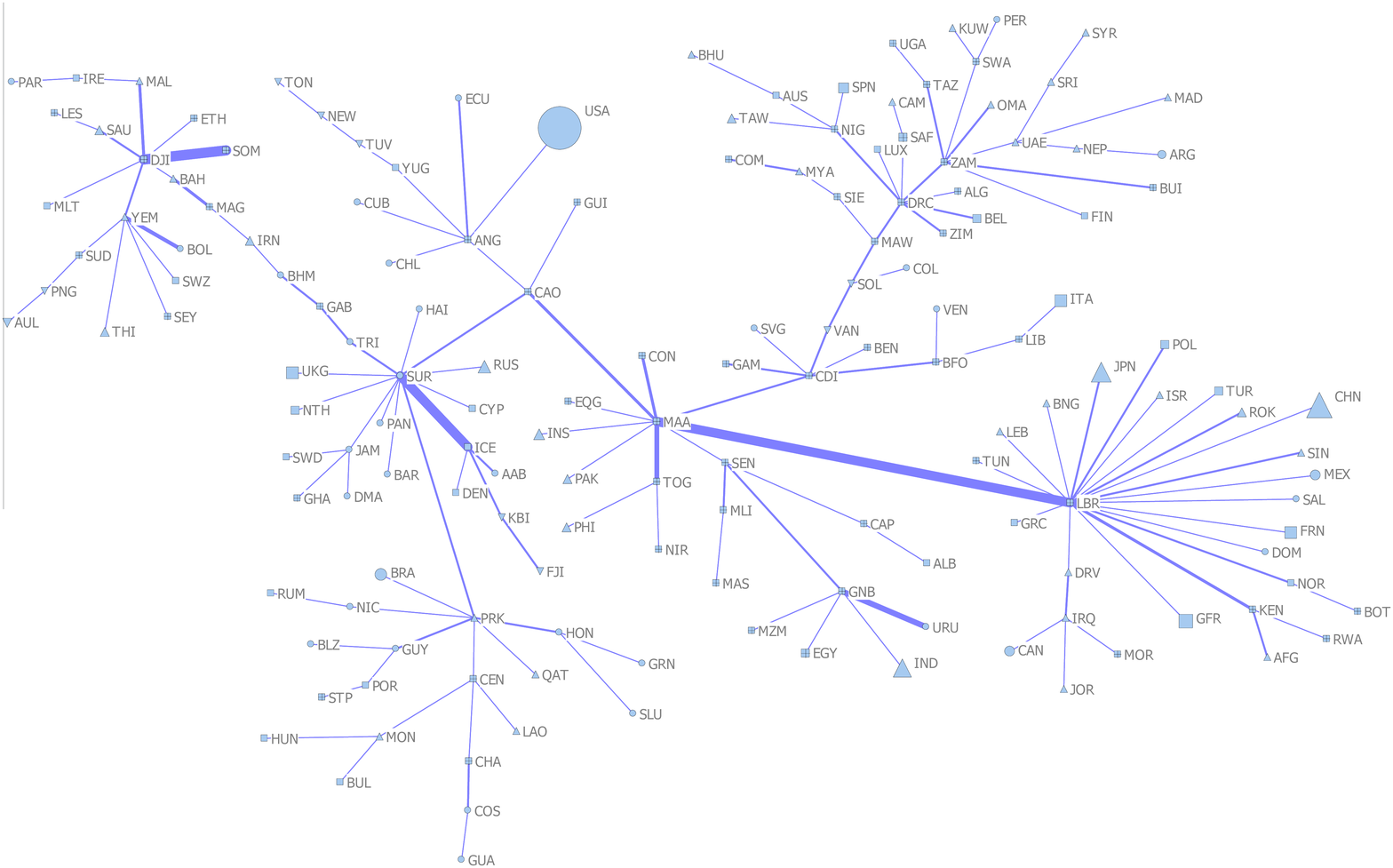}
\caption{\label{Fig:ITN_mst2} Complete minimal spanning tree (MST)
of the residual weighted ITN ($E$). Thickness of a link is
proportional to one minus the weight associated to that link in the
MST returned by the Kruskal's algorithm \citep[][Ch.
23.2]{Cormen01}. Node sizes are proportional to country's GDP. Node
shapes represent the continent which the country belongs to
(Circles: America; Empty Squares: Europe; Upright Triangles: Asia;
Crossed Squares: Africa; Reversed Triangles: Pacific).}
\end{center}
\end{sidewaysfigure}

\end{document}